\documentclass[10pt]{article}
\usepackage{latexsym}

\usepackage[top=1in, bottom=1in, left=1in, right=1in]{geometry}

\usepackage{adjustbox}

\usepackage{amssymb}

\usepackage{graphicx} 
\usepackage{color} 
\usepackage{amsmath, amsthm, amssymb, appendix}
\usepackage{enumerate}
\usepackage{float}
\usepackage{url}
\usepackage{stackrel}
\usepackage{mathrsfs,dsfont}
\usepackage[labelfont=bf]{caption}
\usepackage{subcaption}
\usepackage{wrapfig}
\usepackage{bbm}
\usepackage{bm}
\usepackage{epigraph}
\usepackage[colorinlistoftodos, shadow]{todonotes} 
\usepackage{multirow}
\usepackage{hyperref}
\usepackage{soul}

\usepackage{mdwlist}

\usepackage{pifont}

\newtheorem{theorem}{Theorem}[section]

\newtheorem{lemma}[theorem]{Lemma}
\newtheorem{example}[theorem]{Example}
\newtheorem{remark}[theorem]{Remark}

\theoremstyle{definition}
\newtheorem{assumption}[theorem]{Assumption}

\theoremstyle{definition}
\newtheorem{definition}[theorem]{Definition}
\theoremstyle{definition}

\newcommand{\been}{\begin{enumerate}}
\newcommand{\enen}{\end{enumerate}}
\newcommand{\beit}{\begin{itemize}}
\newcommand{\enit}{\end{itemize}}

\def\GG{\mathcal G}

\def\RR{\mathcal R}

\def\GG{\mathcal{G}}

\def\EE{\mathcal{E}}

\def\ds{\displaystyle}

\def\cre{\color{red}}

\newcommand{\R}{\mathbb{R}}

\newcommand{\Z}{\mathbb{Z}}

\usepackage{tikz}    
\usetikzlibrary{shapes,automata,positioning,arrows,fit}
\usepackage{mathtools}  
\usetikzlibrary{snakes}

\usepackage{caption}
\usepackage{tikz-cd}

\usepackage{tkz-euclide}

\usepackage{tikz-3dplot}

\usepackage{relsize}

 \tikzset{every node/.style={auto}}
 \tikzset{every state/.style={rectangle, minimum size=0pt, draw=none, font=\Large}}
  \tikzset{bend angle=7}
  
  \usepackage{array}
  
  \usepackage{chemfig}

\usepackage{authblk}

\makeatletter
\newcommand{\xrightleftarrows}[2]{
  \mathrel{\mathop{
    \vcenter{\offinterlineskip\m@th
      \ialign{\hfil##\hfil\cr
        \hphantom{$\scriptstyle\mspace{8mu}{#1}\mspace{8mu}$}\cr
        \rightarrowfill\cr
        \vrule height0pt width 2em\cr
        \leftarrowfill\cr
        \hphantom{$\scriptstyle\mspace{8mu}{#2}\mspace{8mu}$}\cr
        \noalign{\kern-0.3ex}
      }
    }
  }\limits^{#1}_{#2}}
}
\makeatother

\begin{document}

\title{Bifunctional enzyme action as a source of robustness \\ in biochemical reaction networks: a novel hypergraph approach}

\author[1*]{Badal Joshi} 
\affil[1]{Department of Mathematics, California State University San Marcos (bjoshi$@$csusm.edu)} 
\author[2]{Tung D. Nguyen} 
\affil[2]{Department of Mathematics, University of California Los Angeles (tungdnguyen$@$math.ucla.edu)}

\date{}

\maketitle

\begin{abstract}
\noindent Substrate modification networks are ubiquitous in living, biochemical systems. A higher-level hypergraph “skeleton" captures key information about which substrates are transformed in the presence of modification-specific enzymes. Many different detailed models can be associated to the same skeleton, however uncertainty related to model fitting increases with the level of detail. 
We show that essential dynamical properties such as existence of positive steady states and concentration robustness can be extracted directly from the skeleton independent of the detailed model. 
The novel formalism of directed hypergraphs is used to prove that bifunctional enzyme action plays a key role in generating robustness. 
Moreover, we use another novel concept of “current'' on a directed hypergraph to establish a link between potentially remote network components. 
Current is an essential notion required for existence of positive steady states, and furthermore,  current-matching combined with bifunctionality generates concentration robustness.

\end{abstract}

\section{Introduction}
Design principles underlying network outputs can reveal insights into the biochemical function, network origins and operating mechanisms. 
Bifunctional enzyme action has been found responsible for output robustness in a series of studies combining mathematical modeling with experimentation \cite{russo1993essential,hsing1998mutations,batchelor2003robustness,shinar2009robustness,dexter2013dimerization}. 
The mechanism relies on a protein that performs opposing kinase and phosphatase activities. 
A robust output is essential for the reliable functioning of thousands of signaling systems \cite{shinar2007input,alon2019introduction,hart2013utility}. 
Building on the previous models, Shinar and Feinberg \cite{shinar2010structural} gave a mathematical theorem with precise, and simple to check, sufficient conditions for concentration robustness. 
These conditions, such as deficiency of one, are strictly mathematical in nature and are not connected with any known biochemical mechanism. 
Moreover, there are at least two limitations to its applicability: (1) the condition of deficiency one restricts the class of networks where the theorem applies, and (2) the theory is silent on existence of positive steady states. 
It is well-known that concentration robustness can arise in networks of any deficiency: low (zero, one) or high (two or more). Most networks in biochemistry have high deficiencies, while the existing theorems apply to low deficiency networks only, see also \cite{joshi2022foundations}.  
In this work, we develop mathematical theory that is closely connected with the structure of biochemical networks and the mechanism of bifunctional enzyme action. 
In doing so we uncover deeper operational principles while simultaneously removing the dependence on deficiency. 
In fact, the deficiency parameter (which does not have a biochemical meaning) is unimportant in this new theory. 
Furthermore, the new theory does not even require the precise reaction network to be known. 
Many different detailed models (which may as yet be experimentally unraveled) may be unified by a simpler description which we call ``skeleton'', and only knowing this will suffice to apply the new results. 
Finally, the new theory also gives precise conditions for existence of positive steady states, a property that is essential for robustness.

The main goal of this paper is to prove the following compact but powerful result and then to apply it to several key biochemical systems to reveal the presence of robustness.

\vspace{5pt}

\begin{center}
\fbox{
\begin{minipage}{35em}
     Suppose that a pair of hyperedges in the skeleton of a substrate modification network is  current-matching and bifunctionally linked. 
    Then the tail of the  e-edge has  concentration robustness in the mass action system of any detailed model. 
\end{minipage}
}
\end{center}

\vspace{5pt}

There are multiple terms in the above statement that will be unfamiliar to readers. 
To give a complete account, we will need to develop these ideas in turn. 
The first ``{\em substrate modification network}'' (see Section \ref{sec:substrate_modification_network}  and Section \ref{sec:submodnetwork}) as a class subsumes most biochemical reaction networks, showing the broad applicability of the theory. 
Next we dive into ``{\em detailed model}'' (see Section \ref{sec:detailed_model}  and Section \ref{sec:detailedmodel}) which captures the inner workings of substrate modifications, and includes the details about the roles of enzymes and intermediate compounds. 
When viewed at a higher level, these inner processes are hidden and several detailed models are unified by a common ``{\em skeleton}''. 
We apply the powerful mathematical formalism of ``{\em hypergraph}'' (instead of the ordinary graph commonly used in reaction network theory) to describe reaction dynamics. 
See Section \ref{sec:skeleton_hypergraph} (and Section \ref{sec:skeleton}) for explanation of skeletons and more generally hypergraphs. 
The notions of ``{\em hypergraph current}'' and ``{\em  current-matching pairs}'' (see Section \ref{sec:current} and Section \ref{sec:current_rep}) are inspired by electric circuit theory and adapted for directed hypergraphs. 
The application of current to directed hypergraphs is novel and, to the best of our knowledge, does not exist in the published literature.  
``{\em Bifunctionality}'' (see Section \ref{sec:bifunctional_enzyme_action} and \cite{shinar2007input}) is a type of action that occurs via the same enzyme in two different places in the network. 
A bifunctional enzyme is produced as a compound in one reaction, and then the compound acts as an enzyme to promote another reaction. The two hyperedges corresponding to these two reactions are the ``{\em c-edge}'' and ``{\em e-edge}'', respectively. 
We refer to the two hyperedges together as a ``{\em bifunctionally linked pair}''.

``{\em Concentration robustness}'' (see Section \ref{sec:ACR}) is where a product of substrate concentrations is invariant across all positive steady states. 
This includes the special case where a single substrate concentration is invariant, referred to as ``{\em absolute concentration robustness}'' in the literature. 
The rough sketch of ideas presented above makes it clear that a detailed exposition is necessary for the pithy but powerful result  in the box to make sense and be readily deployable to biochemical applications. 
This is the purpose of the rest of this paper. 
In  Sections \ref{sec:submodnetwork}-\ref{sec:general}, we give rigorous mathematical definitions of each concept, more examples and explanations, and state and prove rigorous mathematical theorems. 

\begin{figure}
\begin{subfigure}[b]{0.35\textwidth}
\begin{center}
\begin{equation*}
\begin{split}
&X  \xrightleftarrows{}{} X_T \longrightarrow X_p \\
&X_p + Y \xrightleftarrows{}{} X_pY     \longrightarrow  X + Y_p  \\ 
&Y_p + X_T \xrightleftarrows{}{} Y_pX_T \longrightarrow  Y + X_T
\end{split}
\end{equation*}
\caption{EnvZ-OmpR with ATP is a cofactor.}
\end{center}
\end{subfigure}\hfill
\begin{subfigure}[b]{0.25\textwidth}
\begin{center}
\scalebox{0.85}{
\begin{tikzpicture}[baseline={(current bounding box.center)},node distance=2cm, on grid]
    \node (X) {$X$};
    \node (Xp) [left=3cm of X] {$X_p$};
    \node (Y) [below=of Xp] {$Y$};
    \node (Yp) [right=3cm of Y] {$Y_p$};
    \draw[->,line width=1.25] (X) .. controls +(-0.5,1.23) and +(0.5,1.23) .. (Xp);
    \draw[->,line width=1.25] (Xp) .. controls +(0.5,-1.23) and +(-0.5,-1.23) .. (X);
    \draw[->,line width=1.25] (Y) .. controls +(0.5,1.23) and +(-0.5,1.23) .. (Yp);
    \draw[->,line width=1.25] (Yp) .. controls +(-0.5,-1.23) and +(0.5,-1.23) .. (Y);
\end{tikzpicture}} 
\caption{Skeleton for either model.}
\end{center}
\end{subfigure}\hfill
\begin{subfigure}[b]{0.35\textwidth}
\begin{center}
\begin{equation*}
\begin{split}
&X_D \xrightleftarrows{}{} X  \xrightleftarrows{}{} X_T \longrightarrow X_p \\
&X_p + Y \xrightleftarrows{}{} X_pY     \longrightarrow  X + Y_p  \\ 
&Y_p + X_D \xrightleftarrows{}{} Y_pX_D \longrightarrow  Y + X_D
\end{split}
\end{equation*}
\caption{EnvZ-OmpR with ADP is a cofactor.}
\end{center}
\end{subfigure}\hfill

    \caption{Different detailed models of EnvZ-OmpR system have the same skeleton. $X_p$ and $Y_p$ are phosphorylated forms of $X$ and $Y$. $X_D$ and $X_T$ denote the compounds of $X$ with ADP and ATP, respectively. The remaining compounds, such as $Y_pX_T$ is the compound of $Y_p$ and $X_T$, are clear from the notation.}
    \label{fig:envzompr}
\end{figure}
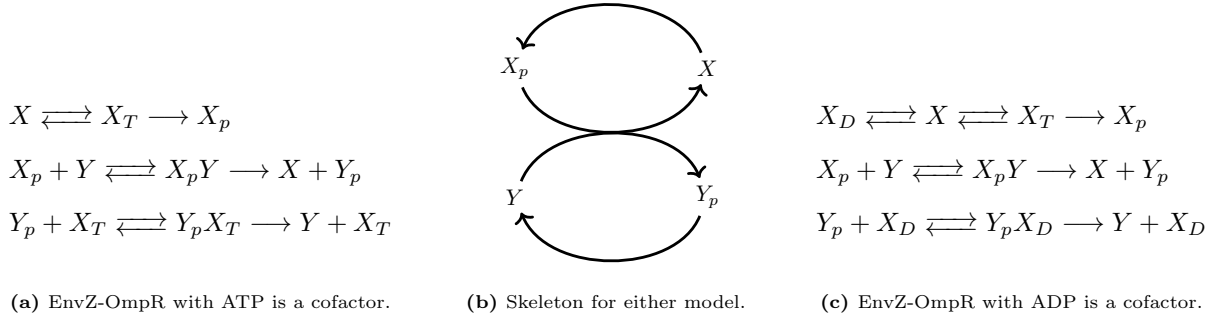

We demonstrate our results on the EnvZ-OmpR system from \cite{shinar2007input}. 
As shown in Figure \ref{fig:envzompr} (a) and (c), two different detailed models (reaction networks) are presented for the EnvZ-OmpR system in \cite{shinar2010structural}. 
Both detailed models have the same substrate skeleton which is a directed hypergraph (Figure \ref{fig:envzompr} (b)). 
The nodes of this hypergraph are the substrates $X$, $X_p$, $Y$ and $Y_p$ (where $X_p$ and $Y_p$ are just phosphorylated forms of $X$ and $Y$). 
The hypergraph has three hyperedges: $X \to X_p$, $Y_p \to Y$ and $\{X_p, Y\} \to \{X, Y_p\}$. 
The crucial distinction between an edge and a hyperedge is that a hyperedge connects a (nonempty) set of nodes to another (nonempty) set of nodes, instead of just one node to another node. 
This particular hypergraph has two hyperedges that are normal edges while one is not. 
For visual depiction, we will often portray a hyperedge as a ``pair of connected split edges'' as we have chosen to do in Figure \ref{fig:envzompr} (b). 
It will turn out that all three hyperedges must have the same current (see Section \ref{sec:current}) because of the structure of the hypergraph. 
Based on the results in this paper, the mere existence of a current then implies that a positive steady state must exist for any detailed mass action model associated to this skeleton. 
Moreover, we will also show that for both detailed models in Figure \ref{fig:envzompr}, the pair of edges $X \to X_p$ and $Y_p \to Y$ is a ``bifunctionally linked pair'' with the former being the  c-edge and the latter being the e-edge. 
The main result of this paper (in the box above) then implies that the source of the  e-edge in the skeleton, that is $Y_p$, has concentration robustness in any detailed model taken with mass action kinetics.
In particular, this is true of the two detailed models that are shown in Figure \ref{fig:envzompr}. 
While deficiency based results of Shinar and Feinberg in \cite{shinar2010structural} do allow the same conclusion for these two models, our results apply far more broadly and do not depend on deficiency at all. 
In Section \ref{sec:discussion}, we comment further on the generality and broad applicability of our results. 

The mathematically inclined reader can find the precise details of the results and applications in Sections \ref{sec:submodnetwork}-\ref{sec:general}. 
Section \ref{sec:submodnetwork} gives a detailed mathematical description of substrate modification networks, including definitions of skeleton, current and what constitutes a detailed model. 
Section \ref{sec:HMM} gives a proof of the main results related to existence of steady states and robustness for the common Henri-Michaelis-Menten detailed model. 
Section \ref{sec:general} then proves these results for a general detailed model.

\section{Substrate modification networks} \label{sec:substrate_modification_network}

Certain reaction archetypes or motifs appear repeatedly in biochemical reaction networks. 
One common motif type is {\em transformation}, wherein a substrate $S$ is transformed to another substrate $P$, typically depicted as a {\em reaction}  $S \to P$. 
Note that we do not make a distinction between forms of a substrate (for instance, phosphorylated and unphosphorylated forms) and different substrates.
 Biochemical systems with such reactions are ubiquitous and include activation-inactivation cycles \cite{goldbeter1981amplified}, cascades \cite{huang1996ultrasensitivity,ferrell1996tripping}, chemotaxis \cite{barkai1997robustness}, signaling \cite{shinar2007input}, metabolic regulation \cite{martinov2000substrate}, gene regulation -- natural \cite{reinitz1990theoretical} and engineered \cite{gardner2000construction}.

Another common biochemical motif is a {\em transfer} reaction, which involves two substrates on the reactant side and two different substrates on the product side. 
For example, in a phosphate transfer reaction, a phosphate group is transferred from a donor $X_p$ to an acceptor $Y$: 
\begin{align} \label{eq:transfer}
X_p + Y~  \longrightarrow   ~X + Y_p. 
\end{align}

Transformation reaction  $S \to P$ and transfer reaction  $S_1 + S_2 \to P_1 + P_2$ are only two types of composite reactions with the same number of substrates on the reactant and the product side, one-one and two-two respectively. 
We give here a complete list of all motif types that involve at most two substrates on the reactant side and at most two substrates on the product side:  
(a) transformation:  $S \to P$,
(b) transfer:  $S_1 + S_2 \to P_1 + P_2$,
(c) cleaving:  $S \to P_1 + P_2$, 
(d) binding:  $S_1 + S_2 \to P$, 
(e) inflow: $0 \to S$, 
(f) outflow: $S \to 0$, 
(g) sequestration:  $S_1 + S_2 \to 0$, and 
(h) pair production: $0 \to  P_1 + P_2$. 
Sometimes the product or reactant species may not be distinct, resulting in higher stoichiometric coefficients. 
For example, the reaction $S + S \to P$ (or $2S \to P$) is allowed as a binding reaction. 
 
Consider {\em glycolysis}, as an illustration of the ubiquitousness of these reaction motifs.  
The overall reaction is a cleaving reaction: 
\[
\rm{glucose} \longrightarrow  \rm{pyruvate} + \rm{pyruvate}. 
\]
When resolved at a higher detail, the standard depiction of glycolysis has 10 steps \cite{alberts2017molecular} with each step involving multiple reactions. Every reaction within each step is one of the following {\em types}: 
(a) transformation reaction $S \to P$ (steps 2, 5, 8), 
(b) transfer reaction $S_1 + S_2 \to P_1 + P_2$ (steps 1, 3, 6, 7, 10), 
(c) cleaving reaction $S \to P_1 + P_2$ (steps 4, 9), 
(d) binding reaction $S_1 + S_2 \to P$ (step 4). 

The probability of more than two substrates binding or reacting is minimal. 
For most biochemical reactions, a higher spatial or temporal resolution will separate a reaction into ones with at most two substrates on either side. 
Therefore such reactions are our focus. 
However, the mathematical results can be applied to situations with a higher number of substrates.
All the reactions described above are types of {\em substrate modification reactions}, and a collection of such reactions is called a {\em substrate modification network}.

\section{Detailed models of a substrate modification network} \label{sec:detailed_model}

A {\em detailed model} goes further than describing the identity of the initial substrates and the product substrates because it also specifies the nature of the enzyme action and intervening reaction steps. 
A common detailed model of the transformation reaction $S  \to P$ uses the Henri-Michaelis-Menten mechanism \cite{johnson2011original,punekar2018henri}
\begin{align} \label{eq:rxnsteps_main}
S + E~  \xrightleftarrows{}{} ~C \longrightarrow P + E,
\end{align}
where $E$ is an {\em enzyme} that facilitates the reaction while $C$ is an intermediate {\em compound}, often denoted as $SE$ to indicate that it is formed from the bonding of $S$ and $E$. $S$ and $P$ are the substrates which also appear in the overall transformation reaction $S \to P$. 
Two different detailed models for the same reaction are:  
\begin{align} 
S + E~  \xrightleftarrows{}{} ~&C_1 \longrightarrow ~C_2 \longrightarrow P + E. \label{eq:rxnsteps6_main_1}\\
S + E~  \xrightleftarrows{}{} ~&C_1 \longrightarrow  C_2 ~  \xrightleftarrows{}{} ~ P + E. \label{eq:rxnsteps6_main_2}
\end{align}

Note that there is a modeling hierarchy from coarse-grained (high confidence) overall reaction description to fine-grained (low confidence) detailed model:   
\begin{enumerate}[(i)]
    \item {\bf (coarse-grained/high confidence)} The model for the overall reaction $S \to P$ simply states that the substrate $S$ is transformed into the substrate $P$ in the presence of an enzyme $E$. 
    \item {\bf (fine-grained/low confidence)} The detailed model goes much further. It describes the nature of the interaction between the enzyme and the substrate, the identity of the intermediate compounds, and the precise reaction steps. 
\end{enumerate}
The first description is necessarily a higher confidence model as it requires a simpler experimental/observational setup to establish or verify. 
One only needs to observe that all the initial $S$ is converted into $P$ after some time in the presence of $E$. 
The time course or the nature of the intermediate compounds  is not essential for this level of description. 
In order to build the second description, we require a more careful experimental design, accurate data about the product formation curves and model fitting to find the best model. 
Thus necessarily, the detailed model is a lower confidence model. 
It is likely that the detailed model is either unknown or uncertain with different models appearing simultaneously in published literature. 

Despite the many possibilities, any individual detailed model of a single substrate modification reaction has three species types: {\em substrate}, {\em enzyme}, and {\em compound} -- the types are based on the roles  of the species in that reaction. 
\begin{enumerate}[(i)]
    \item By {\em substrate}, we mean the species that appear in the overall reaction description. Note that we are not making a distinction between different forms of the same substrate and different substrates. 
    \item By {\em enzyme}, we mean any species that is ultimately neither consumed nor produced in the overall reaction, but whose presence is required for the reaction to proceed. 
    \item By {\em compound}, we mean a molecule formed from the union of a substrate and an enzyme, or from the union of two substrates. 
\end{enumerate}
It is important to note that the species type depends on the reaction. One species can be a compound in one reaction while it is an enzyme or a substrate in another reaction. 
In fact, we will see later that a dual role is essential for bifunctional enzyme action.

The precise set of conditions for an allowed detailed model is given in \ref{sec:detailedmodel} along with many more examples.
In order to distinguish modeling hierarchies, we refer to $S \to P$ as a {\em composite reaction}; the individual reaction networks, such as the ones in \eqref{eq:rxnsteps_main}-\eqref{eq:rxnsteps6_main_2}, as its  {\em detailed model}, and the reactions in each detailed model as {\em reaction steps} for that detailed model.

\section{Skeleton of a substrate modification network} \label{sec:skeleton_hypergraph}

A skeleton of a substrate modification network is a {\em hypergraph} which tracks the substrate modifications while ignoring details about the enzymes and substrate-enzyme compounds. 
Our view is that the skeleton is known even in situations where the detailed model is either unknown or uncertain. The reason is that the reactions appearing in the skeleton simply account for what happens to the substrates at the conclusion of the reaction. 
The results in this paper ultimately only depend on this coarse-grained, high-confidence skeleton model.

We first define a hypergraph as an abstract object after which we proceed to define a skeleton as a certain directed hypergraph. 
A (directed) hypergraph extends the notion of a (directed) graph -- as such it has nodes and (directed) hyperedges, see Section \ref{sec:skeleton} for the precise definition.  
A {\em directed edge} in a {\em graph} connects one tail node to a head node. 
A {\em directed hyperedge} in a {\em hypergraph} connects a nonempty set of tail nodes to a nonempty set of head nodes. 
In a hypergraph representation of a substrate modification network, the set of nodes is the set of substrates, and may additionally include the node $0$. 
For every reaction, there is a hyperedge with the reactant substrates acting as the set of tail nodes while the product substrates acting as the set of head nodes. 
Occasionally, a substrate may have stoichiometry that is 2 or higher. 
In such a case, the substrate is `doubled up', so that the set of tail nodes and the set of head nodes should be more accurately viewed as multisets. 
A {\em multiset} is simply an extension of a set where some element might appear multiple times.

We will use the ``{\em split edge}'' representation to draw a hypergraph skeleton. 
When a composite reaction is a transformation reaction $S_1 \to S_2$, it appears unchanged in the skeleton. 
An inflow reaction or outflow reaction also appears unchanged in a skeleton. 
\begin{figure}[h!] 
\begin{center}
\begin{subfigure}[b]{0.19\textwidth}
\scalebox{0.85}{
\begin{tikzpicture}[node distance=2cm, on grid]
    \node (S1) {$S_1$};
    \node (S2) [below=of S1] {$S_2$};
    \node (S3) [right=3cm of S1] {$P_1$};
    \node (S4) [below=of S3] {$P_2$};
    \draw[->,line width=1.25] (S1) .. controls +(0.5,-1.23) and +(-0.5,-1.23) .. (S3);
    \draw[->,line width=1.25] (S2) .. controls +(0.5,1.23) and +(-0.5,1.23) .. (S4);
\end{tikzpicture}}
\caption{Transfer \\ $S_1+S_2 \to P_1+P_2$}
\end{subfigure}
\begin{subfigure}[b]{0.19\textwidth}
\scalebox{0.85}{
\begin{tikzpicture}[node distance=2cm, on grid]
    \node (S1) {$S$};
    \node (S2) [right=1.5cm of S1] {};
    \node (S3) [above right=1.75cm of S2] {$P_1$};
    \node (S4) [below right=1.75cm of S2] {$P_2$};
    \draw[-,line width=1.25] (S1) .. controls +(0.5,0) and +(0,0) .. (S2);
    \draw[->,line width=1.25] (S2) .. controls +(0,0) and +(0,-1) ..(S3);
    \draw[->,line width=1.25] (S2) .. controls +(0,0) and +(0,1) ..(S4);
\end{tikzpicture}}
\caption{Cleaving \\ $S \to P_1+P_2$}
\end{subfigure}
\begin{subfigure}[b]{0.19\textwidth}
\scalebox{0.85}{
\begin{tikzpicture}[node distance=2cm, on grid]
    \node (S1) {$P$};
    \node (S2) [left=1.5cm of S1] {};
    \node (S3) [above left=1.75cm of S2] {$S_1$};
    \node (S4) [below left=1.75cm of S2] {$S_2$};
    \draw[->,line width=1.25] (S2) .. controls +(0,0) and +(-0.5,0) .. (S1);
    \draw[-,line width=1.25] (S2) .. controls +(0,0) and +(0,-1) ..(S3);
    \draw[-,line width=1.25] (S2) .. controls +(0,0) and +(0,1) ..(S4);
\end{tikzpicture}}
\caption{Binding \\ $S_1+S_2 \to P$}
\end{subfigure}
\begin{subfigure}[b]{0.19\textwidth}
\scalebox{0.85}{
\begin{tikzpicture}[node distance=2cm, on grid]
    \node (S1) {$0$};
    \node (S2) [right=1.5cm of S1] {};
    \node (S3) [above right=1.75cm of S2] {$P_1$};
    \node (S4) [below right=1.75cm of S2] {$P_2$};
    \draw[-,line width=1.25] (S1) .. controls +(0.5,0) and +(0,0) .. (S2);
    \draw[->,line width=1.25] (S2) .. controls +(0,0) and +(0,-1) ..(S3);
    \draw[->,line width=1.25] (S2) .. controls +(0,0) and +(0,1) ..(S4);
\end{tikzpicture}}
\caption{Pair production \\ $0 \to P_1+P_2$}
\end{subfigure}
\begin{subfigure}[b]{0.19\textwidth}
\scalebox{0.85}{
\begin{tikzpicture}[node distance=2cm, on grid]
    \node (S1) {$0$};
    \node (S2) [left=1.5cm of S1] {};
    \node (S3) [above left=1.75cm of S2] {$S_1$};
    \node (S4) [below left=1.75cm of S2] {$S_2$};
    \draw[->,line width=1.25] (S2) .. controls +(0,0) and +(-0.5,0) .. (S1);
    \draw[-,line width=1.25] (S2) .. controls +(0,0) and +(0,-1) ..(S3);
    \draw[-,line width=1.25] (S2) .. controls +(0,0) and +(0,1) ..(S4);
\end{tikzpicture}}
\caption{Sequestration \\ $S_1+S_2\to 0$}
\end{subfigure}
\caption{A hyperedge (shown as a pair of split edges) for common reactions with at most two substrates on each side.}
\label{fig:splitedges}
\end{center}
\end{figure}
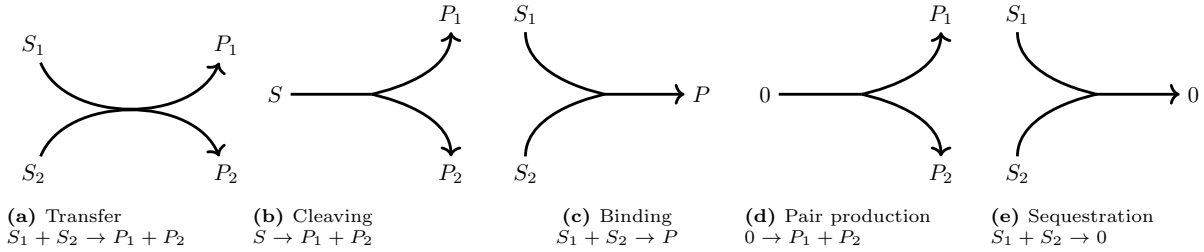
For the reaction $S_1 + S_2 \to P_1 + P_2$ there is a single hyperedge $(\{S_1,S_2\}, \{P_1, P_2\})$, that joins the tail node set $\{S_1,S_2\}$ to the head node set $\{P_1,P_2\}$. 
In the split edge depiction, shown in Figure \ref{fig:splitedges} (a), this appears as a pair of ordinary edges $S_1 \to P_1$ and $S_2 \to P_2$ joined in the middle. However, this is drawn as such only for visual convenience, the choice to visually connect $P_1$ with $S_1$ and $P_2$ with $S_2$ is not significant, and could be reversed with no substantive change. 
 When it is not possible to depict split edges joined in the middle, we use the same color on split edges to indicate that they are part of the same hyperedge.
 Split edge representation is shown in Figure \ref{fig:splitedges}  
for transfer (a), cleaving (b), binding (c), pair production (d), and sequestration (e). 
Note the $0$ acting as a node in some cases. 

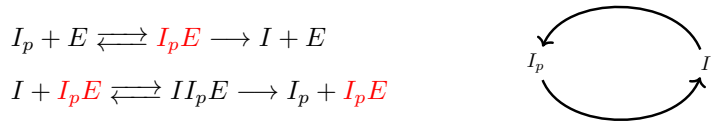
\begin{figure}[h!]
    \centering
\begin{minipage}[c]{0.5\textwidth}
\begin{equation*}
\begin{split}
&I_p + E  \xrightleftarrows{}{} {\cre I_pE} \longrightarrow I + E \\
&I + {\cre I_pE} \xrightleftarrows{}{} II_pE     \longrightarrow  I_p + {\cre I_pE}  
\end{split}
\end{equation*}
\end{minipage}
\begin{minipage}[t]{0.3\textwidth}
\scalebox{0.75}{\begin{tikzpicture}[baseline={(current bounding box.center)},node distance=2cm, on grid]
    \node (X) {$I$};
    \node (Xp) [left=3cm of X] {$I_p$};
    \draw[->,line width=1.25] (X) .. controls +(-0.5,1.23) and +(0.5,1.23) .. (Xp);
    \draw[->,line width=1.25] (Xp) .. controls +(0.5,-1.23) and +(-0.5,-1.23) .. (X);
\end{tikzpicture}}
\end{minipage}
    \caption{The IDHKP-IDH bacterial signaling system has a skeleton with two nodes and two edges which are  current-matching. The compound $I_pE$ produced in the first reaction acts as an enzyme in the second reaction. We highlight this dual role by showing $I_pE$ in red.} 
    \label{fig:idhkpidh}
\end{figure}

Consider the detailed model for the IDHKP-IDH bacterial signaling system with a bifunctional enzyme from \cite{shinar2010structural}, shown on the left in Figure \ref{fig:idhkpidh}. 
When we remove the enzymes  ($E$, $I_pE$) and the intermediate compounds  ($I_pE$, $II_pE$) from consideration, we get the skeleton (right of Figure \ref{fig:idhkpidh}). 
In this case, the skeleton is an ordinary graph that connects the substrates $I$ and $I_p$ with a reversible edge. 

Two different detailed models for the EnvZ-OmpR system have been proposed in 
\cite{shinar2010structural}. In one model, ATP is the cofactor for phospho-OmpR dephosphorylation, while in the second model ADP plays the role of the cofactor. These two detailed models have the same skeleton as shown in Figure \ref{fig:envzompr}.

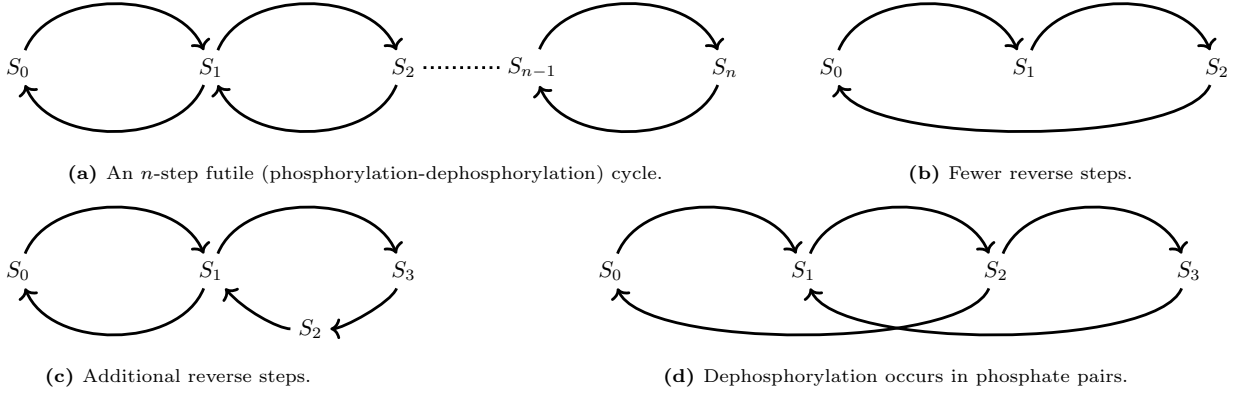
\begin{figure}[h!]
\begin{subfigure}[b]{0.6\textwidth}
\begin{center}
\scalebox{0.85}{
\begin{tikzpicture}[baseline={(current bounding box.center)},node distance=2cm, on grid]
    \node (S0)  {$S_0$};
    \node (S1) [right=3cm of S0] {$S_1$};
    \node (S2) [right=3cm of S1] {$S_2$};
    \node (Sn-1) [right=2cm of S2] {$S_{n-1}$};
    \node (Sn) [right=3cm of Sn-1] {$S_n$};
    \draw[->,line width=1.25] (S0) .. controls +(0.5,1.23) and +(-0.5,1.23) .. (S1);
    \draw[->,line width=1.25] (S1) .. controls +(-0.5,-1.23) and +(0.5,-1.23) .. (S0);
    \draw[->,line width=1.25] (S1) .. controls +(0.5,1.23) and +(-0.5,1.23) .. (S2);
    \draw[->,line width=1.25] (S2) .. controls +(-0.5,-1.23) and +(0.5,-1.23) .. (S1);
    \draw[-,dotted,line width=1.25] (S2) .. controls +(1,0) and +(-1,0) .. (Sn-1);
    \draw[->,line width=1.25] (Sn-1) .. controls +(0.5,1.23) and +(-0.5,1.23) .. (Sn);
    \draw[->,line width=1.25] (Sn) .. controls +(-0.5,-1.23) and +(0.5,-1.23) .. (Sn-1);
\end{tikzpicture}}
\caption{An $n$-step futile (phosphorylation-dephosphorylation) cycle.}
\end{center}
\end{subfigure}\hfill
\begin{subfigure}[b]{0.35\textwidth}
\begin{center}
\scalebox{0.85}{
\begin{tikzpicture}[baseline={(current bounding box.center)},node distance=2cm, on grid]
    \node (S0)  {$S_0$};
    \node (S1) [right=3cm of S0] {$S_1$};
    \node (S2) [right=3cm of S1] {$S_2$};
    \draw[->,line width=1.25] (S0) .. controls +(0.5,1.23) and +(-0.5,1.23) .. (S1);
    \draw[->,line width=1.25] (S2) .. controls +(-0.5,-1.23) and +(0.5,-1.23) .. (S0);
    \draw[->,line width=1.25] (S1) .. controls +(0.5,1.23) and +(-0.5,1.23) .. (S2);
\end{tikzpicture}}
\caption{Fewer reverse steps.}
\end{center}
\end{subfigure}
\begin{subfigure}[b]{0.3\textwidth}
\begin{center}
\scalebox{0.85}{
\begin{tikzpicture}[baseline={(current bounding box.center)},node distance=2cm, on grid]
    \node (S0)  {$S_0$};
    \node (S1) [right=3cm of S0] {$S_1$};
    \node (S3) [right=3cm of S1] {$S_3$};
    \node (dummy) [below=0.9cm of S1] {};
    \node (S2) [right=1.55cm of dummy] {$S_2$};
    \draw[->,line width=1.25] (S0) .. controls +(0.5,1.23) and +(-0.5,1.23) .. (S1);
    \draw[->,line width=1.25] (S1) .. controls +(0.5,1.23) and +(-0.5,1.23) .. (S3);
    \draw[->,line width=1.25] (S3) .. controls +(-0.2,-0.5) and +(0.5,0) .. (S2);
    \draw[->,line width=1.25] (S2) .. controls +(-0.5,0) and +(0.5,-0.6) .. (S1);
    \draw[->,line width=1.25] (S1) .. controls +(-0.5,-1.23) and +(0.5,-1.23) .. (S0);
\end{tikzpicture}}
\caption{Additional reverse steps.}
\end{center}
\end{subfigure}\hfill
\begin{subfigure}[b]{0.55\textwidth}
\begin{center}
\scalebox{0.85}{
\begin{tikzpicture}[baseline={(current bounding box.center)},node distance=2cm, on grid]
    \node (S0)  {$S_0$};
    \node (S1) [right=3cm of S0] {$S_1$};
    \node (S2) [right=3cm of S1] {$S_2$};
    \node (S3) [right=3cm of S2] {$S_3$};

    \draw[->,line width=1.25] (S0) .. controls +(0.5,1.23) and +(-0.5,1.23) .. (S1);
    \draw[->,line width=1.25] (S1) .. controls +(0.5,1.23) and +(-0.5,1.23) .. (S2);
    \draw[->,line width=1.25] (S2) .. controls +(0.5,1.23) and +(-0.5,1.23) .. (S3);
    \draw[->,line width=1.25] (S3) .. controls +(-0.5,-1.23) and +(0.5,-1.23) .. (S1);
    \draw[->,line width=1.25] (S2) .. controls +(-0.5,-1.23) and +(0.5,-1.23) .. (S0);
\end{tikzpicture}}
\caption{ Dephosphorylation occurs in phosphate pairs.} 
\end{center}
\end{subfigure}
    \caption{Skeletons depicting an $n$-step futile cycle and other variations.}
    \label{fig:futilecycle}
\end{figure}

The $n$-step futile cycle \cite{ThomGuna09,WangSontag} has $n+1$ distinct phosphoforms or substrates $S_0, S_1, \ldots, S_n$ that are sequentially phosphorylated and sequentially  dephosphorylated, as shown in Figure \ref{fig:futilecycle} (a). 
Skeletons with different topologies arise from non-sequential phosphorylation or dephosphorylation. 
The three cases presented in Figure \ref{fig:futilecycle} differ in that dephosphorylation when compared with phosphorylation,  either (i) requires fewer steps  in Figure 4 (b), (ii) requires additional steps in Figure 4 (c), or (iii) dephosphorylation occurs in pairs of phosphate groups in Figure 4 (d).  

To wrap up this section, we include a brief discussion on extracting a skeleton from a  detailed model. A detailed discussion on it can be the focus of future papers.
In biochemistry, most detailed models of a composite reaction have a  {\em reactant-product block structure}. In such cases, a skeleton can be obtained unambiguously by removing all intermediate compounds and enzyme. We leave the technical details in Section \ref{sec:detailedmodel}, but roughly speaking, the reactant-product block structure assumes that each detailed model of a composite reaction has
\begin{itemize}
    \item A {\em reactant block} that is made up of a ``reactant substrate $+$ enzyme'' complex and the remaining complexes are its compounds. 
    \item A {\em product block} that is made up of a ``product substrate $+$ enzyme'' complex and the remaining complexes are its compounds.   
    \item A single directed edge going from the reactant block to the product block.
\end{itemize}
To illustrate, we show several examples of detailed models below, where the block structure is evident and therefore we can extract the hyperedge.   
For each detailed model, the reactant block is enclosed in a blue box and the product block is enclosed in a red box. 
The hyperedge for the overall reaction is $S\to S^*$ for all the detailed models shown here, and the enzyme is $E$. 

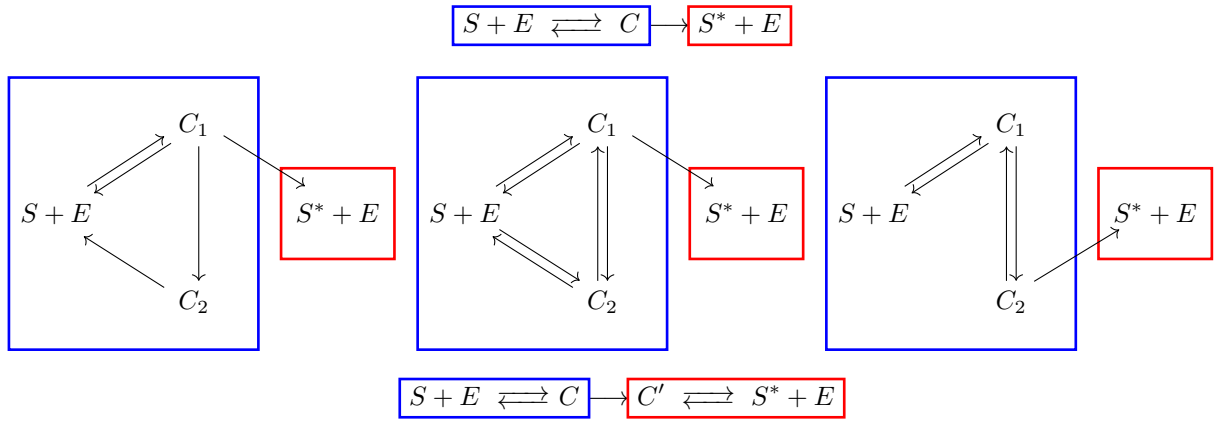
\def\tikzmark#1{\tikz[remember picture,overlay]\node[inner ysep=1pt,anchor=base](#1){\strut};}

\begin{equation*}
\tikzmark{A} S + E~  \xrightleftarrows{}{} ~C \tikzmark{B} \longrightarrow \tikzmark{C} S^* + E \tikzmark{D}
\tikz[remember picture,overlay]\draw[draw=blue,inner xsep=3mm, inner ysep=2mm,line width=1](A.south west)rectangle(B.north east);
\tikz[remember picture,overlay]\draw[draw=red,inner xsep=3mm, inner ysep=2mm,line width=1](C.south west)rectangle(D.north east);
\end{equation*}

\begin{equation*} 
\begin{aligned}
& 
\tikz[overlay]{
    \draw[draw=blue,inner xsep=3mm, inner ysep=2mm,line width=1] (0,-1.8) rectangle (3.3,1.8);
    \draw[draw=red,inner xsep=3mm, inner ysep=2mm,line width=1] (3.6,-.6) rectangle (5.1,.6);
}
\begin{tikzcd}
&  C_1 \arrow[ld,xshift=0.4ex] \arrow[dd,xshift=0.425ex] \arrow[rd,xshift=0.2ex,yshift=0.6ex] \\
S+E \arrow[ru,xshift=-0.2ex,yshift=0.6ex]  &&  S^*+E   \\ 
&  C_2 \arrow[lu,xshift=-0.2ex,yshift=-0.2ex] 
\end{tikzcd} 
& 
\tikz[overlay]{
    \draw[draw=blue,inner xsep=3mm, inner ysep=2mm,line width=1] (0,-1.8) rectangle (3.3,1.8);
    \draw[draw=red,inner xsep=3mm, inner ysep=2mm,line width=1] (3.6,-.6) rectangle (5.1,.6);
}
\begin{tikzcd}
&  C_1 \arrow[ld,xshift=0.4ex] \arrow[dd,xshift=0.425ex] \arrow[rd,xshift=0.2ex,yshift=0.6ex] \\
S+E \arrow[ru,xshift=-0.2ex,yshift=0.6ex] \arrow[rd,xshift=0.4ex,yshift=0.4ex] &&  S^*+E   \\ 
&  C_2 \arrow[lu,xshift=-0.2ex,yshift=-0.2ex] \arrow[uu,xshift=-0.425ex]  
\end{tikzcd} 
& \quad
\tikz[overlay]{
    \draw[draw=blue,inner xsep=3mm, inner ysep=2mm,line width=1] (0,-1.8) rectangle (3.3,1.8);
    \draw[draw=red,inner xsep=3mm, inner ysep=2mm,line width=1] (3.6,-.6) rectangle (5.1,.6);
}
\begin{tikzcd}
&  C_1 \arrow[ld,xshift=0.4ex] \arrow[dd,xshift=0.425ex]  \\
S+E \arrow[ru,xshift=-0.2ex,yshift=0.6ex]  &&  S^*+E   \\ 
&  C_2  \arrow[uu,xshift=-0.425ex] \arrow[ru,xshift=-0.4ex,yshift=0.4ex] 
\end{tikzcd}
\end{aligned}
\end{equation*}

\begin{align*} 
\tikzmark{A} S + E~  \xrightleftarrows{}{} ~&C \tikzmark{B}  \longrightarrow  \tikzmark{C} C' ~  \xrightleftarrows{}{} ~ S^* + E \tikzmark{D} 
\tikz[remember picture,overlay]\draw[draw=blue,inner xsep=3mm, inner ysep=2mm,line width=1](A.south west)rectangle(B.north east);
\tikz[remember picture,overlay]\draw[draw=red,inner xsep=3mm, inner ysep=2mm,line width=1](C.south west)rectangle(D.north east);
\end{align*}

Furthermore, as stated in the previous section, we allow the possibility that a species has a dual role, for instance as both an enzyme and a substrate or as both an intermediary compound and an enzyme. 
When this happens, this species will appear as a node in the hypergraph only in a reaction hyperedge where it has the role of a substrate. 
For instance, when the species has the role of an enzyme and a substrate, it will appear in the hyperedge where it is a substrate but not where it is an enzyme. 
When the species has a role of a compound and an enzyme, it will not appear in the substrate skeleton in any of the two places. However, the pair of hyperedges are considered bifunctionally linked through this compound-enzyme dual role, as explained in Section \ref{sec:bifunctional_enzyme_action}.

\section{Hyperedge current} \label{sec:current}

In a {\em current assignment}, we associate a non-negative quantity (called the ``hyperedge current'') to each hyperedge in a skeleton, which must satisfy a certain balance property defined now. 
The {\em incoming current} at a node $S$ is defined as the sum of all currents for which $S$ is one of the head nodes. 
Similarly, the {\em outgoing current} at $S$ is the sum of all currents for which $S$ is one of the tail nodes. 
A current assignment must satisfy the balance property that at every node, the incoming current equals the outgoing current. 
This is an extension of Kirchoff's current law to a hypergraph.
In addition, we define a {\em positive} current assignment to be a current assignment that is positive on all hyperedges.

To illustrate the concept of currents, we consider the network in Figure \ref{fig:distinguishable}.  
The edge labels $i_\ast$ indicate the current assigned to that edge.  
By the definition of currents (balance property at every node), in {\em any} current assignment, the currents must satisfy: 
\begin{equation} \label{eq:current_assignment_example}
\begin{aligned} 
i_1 &= i_2 \mbox{ (balance at $X_0$ or $X_p$)},  \\ 
i_3 &= i_2 + i_4 \mbox{ (balance at $Y_{00}$)}, \\
i_3 &= i_2 + i_7 \mbox{ (balance at $Y_{p0}$)}, \\
i_5 &= i_6 + i_7 \mbox{ (balance at $Y_{pp}$)},  \\
i_5 &= i_4 + i_6 \mbox{ (balance at $Y_{0p}$)}. 
\end{aligned}
\end{equation}

For additional clarification on the definition, we consider an example which has stoichiometry of 2. For the network in Figure \ref{fig:indistinguishable}, in any current assignment, the currents must satisfy: 
\begin{align*}
       i_1&=i_2\mbox{ (balance at $X$ or $X_p$)},\\
        i_2&=i_3 \mbox{ (balance at $Y_{pp}$)},\\
          2i&_3=i_2+i_1 \mbox{ (balance at $Y_{p}$)},\\
          i_3&=i_1 \mbox{ (balance at $Y$)}.
    \end{align*}

The concept of currents is useful in various situations. For instance, we use it to establish conditions for existence of a positive steady states (see Theorem \ref{thm:consistency}). In the next subsection, we introduce a current-related concept that allows us to state our main results on robustness in Section \ref{sec:results}. 
\subsection{Current-matching pair} \label{sec:current-matching}

Two hyperedges are {\em current-matching}, or alternatively they form a {\em current-matching pair},  if a positive current assignment exists, and if the two hyperedges have the same current for any positive current assignment.  

As an example of current-matching pairs, we look at the three simple motifs appearing in Figures \ref{fig:idhkpidh}, \ref{fig:envzompr} (b), \ref{fig:futilecycle} (b). We reproduce these in Figure \ref{fig:current-balance} with the node identity removed, since this is irrelevant for current-matching. 
As can be easily checked, in each motif all hyperedges have the same current in any current assignment, thus any pair of hyperedges is a current-matching pair.

\begin{figure}
    \centering
    \scalebox{0.85}{
    \begin{tikzpicture}[baseline={(current bounding box.center)},node distance=2cm, on grid]
    \node (X) {$\bullet$};
    \node (Xp) [left=3cm of X] {$\bullet$};
    \draw[->,line width=1.25] (X) .. controls +(-0.5,1.23) and +(0.5,1.23) .. (Xp);
    \draw[->,line width=1.25] (Xp) .. controls +(0.5,-1.23) and +(-0.5,-1.23) .. (X);
\end{tikzpicture} }
\qquad
\scalebox{0.85}{
\begin{tikzpicture}[baseline={(current bounding box.center)},node distance=2cm, on grid]
    \node (X) {$\bullet$};
    \node (Xp) [left=3cm of X] {$\bullet$};
    \node (Y) [below=of Xp] {$\bullet$};
    \node (Yp) [right=3cm of Y] {$\bullet$};
    \draw[->,line width=1.25] (X) .. controls +(-0.5,1.23) and +(0.5,1.23) .. (Xp);
    \draw[->,line width=1.25] (Xp) .. controls +(0.5,-1.23) and +(-0.5,-1.23) .. (X);
    \draw[->,line width=1.25] (Y) .. controls +(0.5,1.23) and +(-0.5,1.23) .. (Yp);
    \draw[->,line width=1.25] (Yp) .. controls +(-0.5,-1.23) and +(0.5,-1.23) .. (Y);
\end{tikzpicture}}
\qquad
\scalebox{0.85}{
\begin{tikzpicture}[baseline={(current bounding box.center)},node distance=2cm, on grid]
    \node (S0)  {$\bullet$};
    \node (S1) [right=3cm of S0] {$\bullet$};
    \node (S2) [right=3cm of S1] {$\bullet$};
    \draw[->,line width=1.25] (S0) .. controls +(0.5,1.23) and +(-0.5,1.23) .. (S1);
    \draw[->,line width=1.25] (S2) .. controls +(-0.5,-1.23) and +(0.5,-1.23) .. (S0);
    \draw[->,line width=1.25] (S1) .. controls +(0.5,1.23) and +(-0.5,1.23) .. (S2);
\end{tikzpicture}}
\caption{Three motifs from Figures \ref{fig:idhkpidh}, \ref{fig:envzompr} (b), \ref{fig:futilecycle} (b). In all three motifs, all pairs of hyperedges are  current-matching.}  
    \label{fig:current-balance}
\end{figure}
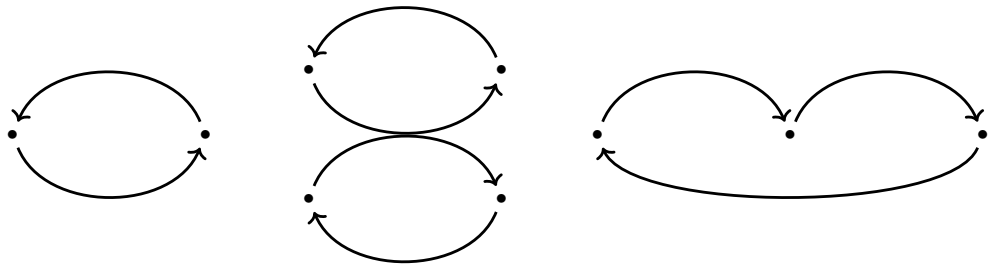

\begin{figure}[h!]
\begin{center}
\scalebox{0.85}{
\begin{tikzpicture}[baseline={(current bounding box.center)}, scale=1]

   \node[state] (Xp)  at (0,1.9)  {$X_p$};
   \node[state] (X0)  at (3.5,1.9)  {$X_0$};
   \node[state] (Y00)  at (0,0)  {$Y_{00}$};
   \node[state] (Y0p)  at (0,-3)  {$Y_{0p}$};
   \node[state] (Ypp)  at (3.5,-3)  {$Y_{pp}$};
   \node[state] (Yp0)  at (3.5,0)  {$Y_{p0}$};

    \path[thick,->]
   (X0) edge[->,violet,bend right =45,line width=1.5] node {$i_1$} (Xp)
   (Xp) edge[->,red,bend right =45,line width=1.5] node {$i_2$} (X0)
   (Y00) edge[->,blue,line width=1.5] node {$i_4$} (Y0p)
   (Y0p) edge[->,olive,bend left =45,line width=1.5] node {$i_5$} (Ypp)
   (Ypp) edge[->,black,line width=1.5] node {$i_7$} (Yp0)
   (Yp0) edge[->,yellow!80!black,bend left =45,line width=1.5] node {$i_3$} (Y00)
   (Y00) edge[->,red,bend left =45,line width=1.5] node {} (Yp0)
   (Ypp) edge[->,green,bend left =45,line width=1.5] node {$i_6$} (Y0p);           
  \end{tikzpicture}}
\caption{{\em Interacting substrates with distinguishable phosphorylation sites.} Substrate $X$ has a single phosphorylation site, while $Y$ has two distinguishable sites.  In any current assignment, we have $i_1 = i_2$ and $i_4 = i_7$.}
\label{fig:distinguishable}
\end{center}
\end{figure}
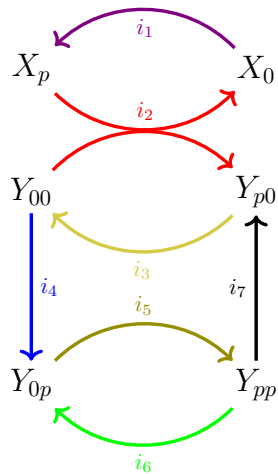

\begin{figure}[h!]
\begin{center}
  \scalebox{0.85}{
  \begin{tikzpicture}[baseline={(current bounding box.center)}, scale=1]

   \node[state] (X)  at (0,1.9)  {$X$};
   \node[state] (Xp)  at (3.5,1.9)  {$X_p$};
   \node[state] (Yp)  at (0,0)  {$Y_p$};
    \node[state] (Y)  at (3.5,0)  {$Y$};
   \node[state] (Z)  at (0,-1.9)  {$Z$};
   \node[state] (Zp)  at (3.5,-1.9)  {$Z_p$};

    \path[thick,->]
   (X) edge[->,red,bend right =45,line width=1.5] node {$i_2$} (Xp)
    (Xp) edge[->,violet,bend right =45,line width=1.5] node {$i_1$} (X)
    (Yp) edge[->,red,bend left =45,line width=1.5] node {} (Y)
    (Y) edge[->,olive,bend left =45,line width=1.5] node {$i_3$} (Yp)
    (Zp) edge[->,olive,bend right =45,line width=1.5] node {} (Z)
    (Z) edge[->,violet,bend right =45,line width=1.5] node {} (Zp)
   ;           
  \end{tikzpicture}}
\caption{ {\em Cyclic phosphate transfer} between three substrates. The reactions are $X + Y_p \to X_p + Y$, $Y + Z_p \to Y_p + Z$, $Z + X_p \to Z_p + X$. In any current assignment, we have $i_1 = i_2=i_3$.}
\label{fig:cyclic_transfer}
\end{center}
\end{figure}
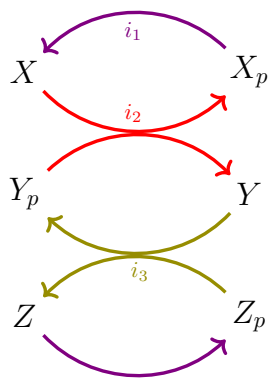

\begin{figure}[h!]
\begin{center}
  \scalebox{0.85}{
  \begin{tikzpicture}[baseline={(current bounding box.center)}, scale=1]
   \node[state] (X)  at (0,1.9)  {$X$};
   \node[state] (Xp)  at (3.5,1.9)  {$X_p$};
   \node[state] (Ypp)  at (0,0)  {$Y_{pp}$};
   \node[state] (Yp)  at (3.5,0)  {$Y_p$};
   \node[circle,fill=yellow!80!black] (Z)  at (3.5,-1.4)  {};
   \node[state] (Y)  at (7.0,0)  {$Y$};

    \path[thick,->]
   (X) edge[->,red,bend right =45,line width=1.5] node {$i_2$} (Xp)
    (Xp) edge[->,violet,bend right =45,line width=1.5] node {$i_1$} (X)
    (Ypp) edge[->,red,bend left =45,line width=1.5] node {} (Yp)
    (Y) edge[->,violet,bend right =45,line width=1.5] node {} (Yp)
    (Yp) edge[-,olive,bend right=30,line width=1.5] node {} (Z)
    (Yp) edge[-,olive,bend left=30,line width=1.5] node {} (Z)
    (Z) edge[->,olive,bend left =45,line width=1.5] node {$i_3$} (Ypp)
    (Z) edge[->,olive,bend right =45,line width=1.5] node {} (Y)
   ;           
  \end{tikzpicture}}
  \caption{{\em Interacting substrates with indistinguishable phosphorylation sites.} A double edge emerging from $Y_p$ indicates a stoichiometry of $2$ in the reaction $2Y_p \to Y_{pp} + Y$. In any current assignment, we have $i_1 = i_2=i_3$. }
  \label{fig:indistinguishable}
  \end{center}
\end{figure}
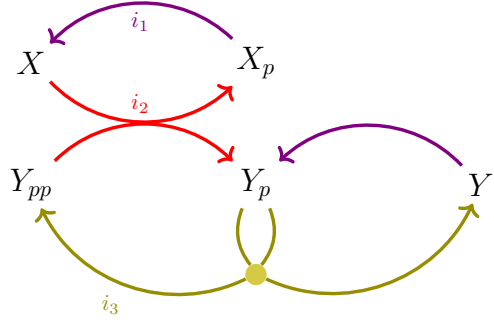

We revisit another example, which is the network given in Figure \ref{fig:distinguishable}. 
Recall that the currents must satisfy \eqref{eq:current_assignment_example}. 
From these equalities, we can obtain  $i_1 = i_2$ and $i_4 = i_7$, which implies there are two  current-matching pairs $\{i_1, i_2\}$ and $\{i_4, i_7\}$. The latter demonstrates that a current-matching pair need not be an adjoining pair in the hypergraph.

\section{Bifunctional enzyme action}
\label{sec:bifunctional_enzyme_action}

Thus far we have focused on substrate modifications and ignored the details about enzymes and intermediate compounds. 
We need to add a little more information to the skeleton -- {\em the identity} of the bifunctional enzyme -- without going all the way to specifying the detailed model.

A {\em bifunctional enzyme} performs different functions in two distinct composite reactions. 
In one reaction, it is produced as a compound, for instance a {\em substrate-enzyme compound} in a transformation reaction or a {\em substrate-substrate compound} in a transfer reaction. 
This compound then acts as an enzyme in another composite reaction. 
In the skeleton, the hyperedge corresponding to the composite reaction in which the bifunctional enzyme is produced  will be referred to as the {\em c-edge}. The hyperedge for the reaction in which it acts as an enzyme will be referred to as the {\em e-edge}.
The two hyperedges collectively (c-edge and e-edge) form a {\em bifunctionally linked pair of hyperedges} in the skeleton.

For the biochemistry and models of bifunctional enzymes, see \cite{shinar2007input,alon2019introduction,hart2013utility,russo1993essential,hsing1998mutations,batchelor2003robustness,shinar2009robustness,dexter2013dimerization}. 
There are two ways in which our terminology differs from the literature: (i) we refer to the compound as bifunctional, not the isolated enzyme, since the compound reappears as an enzyme in another reaction, and (ii) the compound may be a substrate-substrate compound in a transfer reaction that acts as an enzyme in another reaction. There is no isolated enzyme in the latter case.

The IDHKP-IDH bacterial signaling system, shown in Figure \ref{fig:idhkpidh}, has  a bifunctional enzyme $I_pE$. 
The unphosphorylated isocitrate dehydrogenase (IDH) is the active form in the bacterial glyoxylate bypass regulation system, and is denoted by $I$. 
$E$ acts as a phosphatase for the composite reaction $I_p \to I$. 
When bound with $I_p$, the compound $I_pE$ acts as a kinase for $I \to I_p$. 
Thus, $\{I_p \to I, I \to I_p\}$ is a bifunctionally linked pair, with the former a c-edge and the latter an e-edge in the hypergraph.

\section{Absolute concentration robustness} \label{sec:ACR}

A species has absolute concentration robustness (ACR) if its concentration is invariant across all positive steady states. 
A simple example of a substrate modification network with ACR is the IDHKP-IDH system shown in Figure \ref{fig:idhkpidh} and again here with edges labelled with mass action rate constants
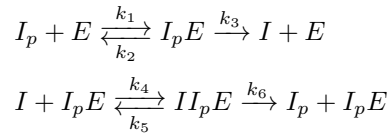
\begin{equation*}
\begin{split}
&I_p + E  \xrightleftarrows{k_1}{k_2} I_pE \xrightarrow{k_3} I + E \\
&I + I_pE \xrightleftarrows{k_4}{k_5} II_pE     \xrightarrow{k_6}  I_p + I_pE  
\end{split}
\end{equation*}
Using the notation $[X]$ for the concentration of any given species $X$, the mass action differential equations for this system are: 
\begin{equation}\label{eq:diffeq_idhkpidh}
\begin{aligned}
\frac{d\left[I_p \right]}{dt}  &= - k_1 \left[I_p \right]\left[E \right]  + k_2 \left[I_pE \right] + k_6 \left[I I_pE \right], \\
\frac{d\left[E \right]}{dt}  &= - k_1 \left[I_p \right]\left[E \right]  + (k_2+k_3) \left[I_pE \right], \\
\frac{d\left[I_p E \right]}{dt}  &=  k_1 \left[I_p \right]\left[E \right]  - (k_2+k_3) \left[I_pE \right] - k_4 \left[I \right]\left[I_pE \right] + (k_5 + k_6) \left[I I_pE \right], \\
\frac{d\left[I \right]}{dt}  &= k_3 \left[I_pE \right] - k_4 \left[I \right]\left[I_pE \right] + k_5 \left[I I_pE \right], \\
\frac{d\left[I I_p E \right]}{dt}  &=   k_4 \left[I \right]\left[I_pE \right] - (k_5 + k_6) \left[I I_pE \right]. 
\end{aligned}
\end{equation}
There are two conserved quantities: (i) the total substrate $\left[ I_p\right] + \left[ I\right] + \left[ I_p E\right] + 2 \left[I I_p E\right]$, and (ii) the total enzyme $\left[ E\right] + \left[ I_pE \right] + \left[ II_pE\right]$. 
One can show with a calculation that as long as the initial total substrate is strictly greater than 
\begin{equation} \label{eq:ACRvalueIDHKPIDH}
   I^* \coloneqq \frac{k_3}{k_4}\left(1+\frac{k_5}{k_6}\right),  
\end{equation}
then there is a positive steady state, and in fact there is only one positive steady state for a fixed value of initial substrate and enzyme (also see \cite{joshi2024bifunctional} for more details on steady states of such systems). 
In particular, this means that there are infinitely many distinct positive steady states, one for each value of initial total substrate and initial total enzyme. 
Every coordinate of the positive steady state, except for $\left[ I \right]$ -- the unphosphorylated IDH, has dependence on these initial values. 
The value of $\left[ I \right]$ at a positive steady state is always $I^*$ (as defined above in \eqref{eq:ACRvalueIDHKPIDH}), a quantity that only depends on the rate constants. 
Therefore, we call $I$ an ACR species and $I^*$ as its ACR value. 

Generalizing the notion of ACR, it is useful to consider {\em concentration robustness of a complex}. 
By {\em complex}, we mean a positive-integer linear combination of species. 
For example, we say that the complex $2X + Y$ has concentration robustness if the product of concentrations, i.e. $\left[ X\right]^2\left[ Y\right]$ is invariant across all positive steady states.

In the next section, we give a simple condition on the skeleton of a substrate modification network that ensures concentration robustness in the mass action system. 

\section{Main results}\label{sec:results}

Our main result is that bifunctional enzyme action produces concentration robustness in the targeted substrate complex. 
Furthermore, the connection between bifunctionality and robustness is mediated by current-matching in the skeleton. 
The main results are summarized below. 
The precise theorems with all the definitions,  mathematical notation, and proofs can be found in Sections \ref{sec:submodnetwork}-\ref{sec:general}.

\vspace{5pt}

\begin{center}

\fbox{
\begin{minipage}{38em}
{\bf A. (Existence of positive steady state).}
Suppose there exists a positive current assignment on the skeleton of a substrate modification network. 
Then the mass action system of any detailed model has a positive steady state for some rate constants. 
\end{minipage}
}

\end{center}

\begin{center}

\fbox{
\begin{minipage}{38em}
{\bf B. (Robustness).}
    Suppose that a pair of hyperedges in the skeleton of a substrate modification network is  current-matching and bifunctionally linked. 
    Then the tail of the  e-edge has  concentration robustness in the mass action system of any detailed model. 
\end{minipage}
}

\end{center}

\begin{center}

\fbox{
\begin{minipage}{38em}
{\bf C. (ACR).}
    Suppose that a pair of hyperedges in the skeleton of a substrate modification network is  current-matching and  bifunctionally linked. 
    If the tail of the  e-edge has only one substrate then this substrate has absolute  concentration robustness in the mass action system of any detailed model. 
\end{minipage}
}

\end{center}

\subsection{Extension to kinetics beyond mass action}
\label{sec:non-massaction}

The dynamics of species concentrations are governed by not only the detailed model but the type of kinetics. 
Commonly used kinetics, besides mass-action, are Michaelis-Menten and more generally, Hill kinetics. 
For these kinetics, the rates of reactions are non-linear and saturating in substrate concentration but linear in enzyme concentration. 
For instance for Michaelis-Menten kinetics (not to be confused with Henri-Michaelis-Menten detailed model), the rate of the composite reaction $S \to P$ in presence of enzyme $E$ is proportional to
\[
\frac{[S][E]}{K_{max} + [S]}. 
\]
where $K_{max}$ is a fixed positive constant. 
All results in this paper extend to such kinetics. 
The only requirement on the kinetics is that the rate functions are proportional to the enzyme concentrations and to the concentration of substrate-enzyme compounds.

\section{Applications to biochemistry}

As a demonstration of the reach of our framework, we will now apply it to some prominent substrate modification networks.
For some of the applications such as IDHKP-IDH and EnvZ-OmpR signaling systems, when a specific detailed model is assumed, the deficiency can be calculated and for some detailed models happens to equal one. In such cases, Shinar-Feinberg theory will also give robustness results that agree with our theory.

However, for most classes of substrate modification networks, the deficiency is either unknown, unknowable (because the detailed model is not known or is uncertain) or known to be greater than 1. In such situations, previous results such as Shinar-Feinberg theory will not apply while our results will still yield robustness properties.

\begin{enumerate}
    \item {\bf (IDHKP-IDH)}: The two composite reactions in the skeleton of the IDHKP-IDH system are clearly current-matching. 
    In particular, clearly a positive current exists which implies the existence of a positive steady state for some mass action rate constants. 
    As noted in the previous section, the two hyperedges are a bifunctionally linked pair and therefore the source of the e-edge, the unphosphorylated IDH $I$, has ACR for any detailed model, including specifically the detailed model in \eqref{fig:idhkpidh}. 
    
    \item {\bf (EnvZ-OmpR)}: Now consider the EnvZ-OmpR ``ping-pong'' system with $X_T$, the $X$-ATP compound as bifunctional enzyme, as shown in Figure \ref{fig:envzompr} (a). 
    One composite reaction is a phosphate transfer reaction, with $X_p$ as donor and $Y$ as acceptor. The reaction is represented by split edges in the skeleton. 
    The phosphorylation of $X$ occurs via $X_T$, a compound of $X$ with an ATP molecule. 
    The reaction actually involves a transfer of a phosphate from ATP to $X$ but ATP and ADP are conventionally excluded since their concentrations are assumed constant due to other processes. 
    The compound $X_T$ has a bifunctional enzyme action in the dephosphorylation of $Y_p$. 
    The tail of the e-edge in the skeleton is $Y_p$ and thus this species has ACR. 

    The scheme in Figure \ref{fig:envzompr} (c) differs only in the bifunctional enzyme, the compound $X$-ADP, denoted by $X_D$ and an extra pair of reactions resulting from formation and disassociation of this compound. The skeleton is unchanged and so is the bifunctionally linked pair, thus leading to the same conclusion of ACR in $Y_p$.

    \item {\bf (Futile cycle with bifunctional enzyme)}: We gave conditions for ACR in the $n$-step futile cycle with a bifunctional enzyme in \cite{joshi2024bifunctional}, while assuming the standard Henri-Michaelis-Menten detailed model.  
    Here we give skeleton-current based conditions which are not only simpler to verify but do not depend on the detailed model and are therefore a broad generalization of the results in \cite{joshi2024bifunctional}.

    It is clear from Figure \ref{fig:futilecycle} (a) that a pair of hyperedges is current-matching if and only if they form a loop. 
    Suppose the loop formed by the pair of reactions $S_j \rightleftarrows S_{j+1}$ where $0 \le j \le n-1$ is a  bifunctionally linked pair. 
    If the edge $S_{j+1} \to S_j$ is the e-edge, i.e. the intermediate compound of $S_j \to S_{j+1}$ is the bifunctional enzyme for the reaction $S_{j+1} \to S_j$, then the futile cycle has ACR in $S_{j+1}$. On the other hand, if $S_j \to S_{j+1}$ is the e-edge then the futile cycle has ACR in $S_j$. 

    \item  {\bf (Futile cycle variant -- phosphate groups attach singly but detach in pairs -- Figure \ref{fig:futilecycle} (d).)}: Refer to the current labeling in Figure \ref{fig:jump_return}. From current balance at the node $S_0$, $i_1=i_5$, and from current balance at $S_3$, $i_3=i_4$. This makes $\{i_1,i_5\}$ and $\{i_3,i_4\}$ current-matching pairs.     
    Current balance at the other two nodes reveals that $i_2 = i_1 + i_3$. For the general case of $n$ phosphorylation sites, current balance at the end nodes $S_0$ and $S_n$ will give two current-matching pairs. 
    We consider 4 cases of bifunctionally linked edges in Figure \ref{fig:jump_return}: (i) $i_1$ is c-edge and $i_5$ is e-edge $\implies$ $S_2$ has ACR, (ii) $i_5$ is c-edge and $i_1$ is e-edge $\implies$ $S_0$ has ACR, (iii) $i_3$ is c-edge and $i_4$ is e-edge $\implies$ $S_3$ has ACR, and (iv) $i_4$ is c-edge and edge $i_3$ is e-edge $\implies$ $S_2$ has ACR. Interestingly, there are two ways in which the species $S_2$ can have ACR -- each of the outgoing edges of $S_2$ is one of a current-matching pair. On the other hand, there is no way that $S_1$ can have ACR -- the only outgoing edge is not one of a current-matching pair.

    \item  {\bf (Interacting substrates with distinguishable phosphorylation sites -- Figure \ref{fig:distinguishable})}: 
The substrate $X$ has one phosphorylation site. 
The unphosphorylated form of $X$ is denoted $X_0$ while the phosphorylated form is denoted by $X_p$. 
The substrate $Y$ has two (distinguishable) phosphorylation sites, and therefore has four distinct phosphoforms: $Y_{00}, Y_{p0}, Y_{0p}$, and $Y_{pp}$.

We have argued in Section \ref{sec:current-matching} that there are two current-matching pairs: $\{i_1, i_2\}$ and $\{i_4, i_7\}$. 
Consider a situation where one of the pairs is bifunctionally linked. 
For the first pair, assuming that $i_2$ is the c-edge and $i_1$ is the e-edge, the source of the e-edge $X_0$ will have absolute concentration robustness in any detailed model. 
On the other hand, if the c-edge and e-edge are reversed, then the complex $X_p + Y_{00}$ will have robustness. In other words, the product of concentrations $x_p y_{00}$ will be invariant across all compatibility classes in any detailed model. 

If the bifunctionally linked pair is $\{i_4, i_7\}$ then either $Y_{00}$ or $Y_{pp}$ can have ACR depending on which node is the tail of the e-edge. 

We mention the interesting possibility of both current-matched pairs being bifunctionally linked and $i_2$ and $i_4$ representing the e-edges in the two pairs. The robustness of $X_p + Y_{00}$ along with ACR in $Y_{00}$ automatically implies ACR in $X_p$.

\item     {\bf (Cyclic phosphate transfer)}:
    Consider a set of $n$ substrates $X_1, \ldots, X_n$ that transfer phosphate in cyclic order, each reaction is: 
\begin{align*}
    X_i + X_{i+1,p} \longrightarrow X_{i,p} + X_{i+1},  \quad (1 \le i \le n), 
\end{align*}
$X_{n+1} = X_1$, and  $X_{i,p}$ is the phosphorylated form of $X_i$. 
We will denote the unphosphorylated form simply by $X_i$. 
The skeleton has $n$ hyperedges, one for each phosphate transfer reaction. 
A schematic diagram for three substrates is presented in Figure \ref{fig:cyclic_transfer}, where we relabeled the species $X_1, X_2, X_3$ as $X, Y, Z$. 

It is simple to verify that any pair of hyperedges is a current-matching pair. This means that if any pair is bifunctionally linked, the tail of the e-edge will have robustness. 
Since the tail has the generic form $X_i + X_{i+1,p}$, the invariant quantity across positive steady states is a product $x_i x_{i+1,p}$.

    \item {\bf (Interacting substrates with indistinguishable phosphorylation sites -- Figure \ref{fig:indistinguishable}}:
    We again consider a situation where a species $X$ has one phosphorylation site while a species $Y$ has two (indistinguishable) phosphorylation sites, so the phosphoforms are $X, X_p, Y, Y_p$, and $Y_{pp}$. 
Three transfer reactions are possible: 
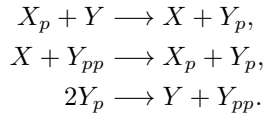
\begin{align*}
    X_p + Y &\longrightarrow X + Y_p, \\
    X + Y_{pp} &\longrightarrow X_p + Y_p, \\
    2Y_p &\longrightarrow Y + Y_{pp}. 
\end{align*}
The last composite reaction produces a hyperedge from the {\em multiset} $\{Y_p, Y_p\}$ to the set $\{Y, Y_{pp}\}$. 

Similar to the previous case, it is easy to verify that any pair of hyperedges is a current-matching pair. 
Concentration robustness follows if any pair is bifunctionally linked.  
An interesting possibility arises when the last hyperedge $2Y_p \to Y + Y_{pp}$ is the e-edge in a bifunctionally linked pair. 
Then the complex $2Y_p$ has robustness from which it follows that $y_p^2$ is invariant across positive steady states, and therefore $y_p$ is also invariant. In other words, the species $Y_p$ has ACR. 

\end{enumerate}

\section{Discussion}
\label{sec:discussion}

Substrate modification networks are ubiquitous in biochemistry. 
There has been tremendous interest in the mathematical study of their dynamics in recent years. 
Many works focus on the analysis of individual networks that are either biochemically or mathematically interesting. 
Often substrate modification networks are simply treated as one possible type of reaction network. 
We take the view here that substrate modification networks, and somewhat more generally enzymatic reaction networks are what distinguish biochemistry from chemistry in general. 
Processes of life require the underlying networks to have special structures, specifically reaction steps involving substrate-enzyme bonding and compound formation, followed with release of enzyme concurrent with transformation of substrate(s). 

We introduce the use of hypergraphs to study the underlying structure of substrate modification networks. 
A focus on the substrates, while hiding the features of enzymes and substrate-enzyme compounds, reveals important structural details which are depicted as the skeleton of the substrate modification network. 
Studying the skeleton has the advantage that one skeleton unifies multiple different detailed models. 
Results established for the skeleton simultaneously apply to all the detailed models. 
Not only does this reveal the underlying mechanism with striking clarity, it has another distinct advantage. 
A skeleton can be experimentally established easily and with much higher confidence than the detailed model. 
Therefore, results about the skeleton are valid with a much higher confidence. 
Furthermore, inspired by electrical networks, we  introduce the concept of {\em hypergraph current} on a skeleton. 
The notion of current is  novel to both areas -- the mathematical study of hypergraphs and that of reaction networks. 
Hypergraph current provides a connection between different edges or hyperedges, possibly distantly located within the substrate skeleton. 

In this paper, we hone into a particular mechanism -- that of bifunctional enzyme action  -- that gives rise to absolute concentration robustness (ACR). 
Previous mathematical literature on ACR has focused on deficiency theory. 
There are two clear drawbacks of deficiency-based theory as applied to ACR -- (i) the theory only applies to networks with deficiency one and deficiency zero, (ii) deficiency by itself has no interpretation which connects it to a biochemical function or feature. 
By zooming into the connection with bifunctional enzymes, we connect a mathematical condition to a biochemical function. 
Moreover, the mathematical framework of substrate skeleton, hypergraph current, and bifunctional enzyme action is completely oblivious of deficiency. 
It applies to networks of any deficiency whatsoever. 
Moreover, since the detailed reaction steps are unimportant, the results are extremely robust against modeling inaccuracies or discrepancies. 
A detailed model is a choice of molecular dynamics that is suggested by limited experimental dynamical data. 
Many different detailed models might fit the data within a certain tolerance. 
Our results suggest that the property of concentration robustness and existence of positive steady states does not depend critically on these details, and continues to hold even if the detailed model is not firmly established.

\renewcommand{\thesection}{S\arabic{section}}

\section*{Supporting Mathematical Theory}

In the following sections, we provide precise mathematical definitions, notations, theorems which support our three main results on existence of positive steady state, concentration robustness (of a complex), and ACR, together with their corresponding proofs. 

\section{Substrate modification networks}\label{sec:submodnetwork}

The main purpose of this section is to recall relevant definitions arising from substrate modification networks and set up notations which are used in later sections. The section is structured as follows. In Section \ref{sec:skeleton} and \ref{sec:current_rep}, we recall skeleton and hypergraph current. Next, in Section \ref{sec:reactionnetwork}, we present some necessary terminology from reaction network theory in order to describe different types of detailed models in Section \ref{sec:detailedmodel}. Lastly, in Section \ref{sec:enzyme}, we introduce convenient notation to specify the enzyme acting in each composite reaction.

\subsection{Skeleton}\label{sec:skeleton}

We consider a substrate modification network with $n$ substrates $S_1, S_2,\dots, S_n$. For the sake of convenience and consistency, we also introduce the ``null substrate" $S_\emptyset$ to represent the zero complex which appears in inflow and outflow reactions. For example, the inflow reaction $0\to S_1$ will be written interchangeably as $S_\emptyset\to S_1.$ 

The \textit{skeleton} of the substrate modification network is a (directed) hypergraph, which is denoted by $G=(V,R)$. 
The set of vertices is $V=\{S_\emptyset, S_1, \dots, S_n\}$ ($S_\emptyset$ is omitted if the network does not contain the $0$ complex). The set of hyperedges represents the composite reactions in the substrate modification network. For each hyperedge $\ell\in R$, let $T_\ell$ and $H_\ell$ be the set of indices of the tail vertices and head vertices. Each hyperedge $\ell$ represents a composite reaction 
\[
\sum_{i\in T_\ell} S_i \to \sum_{i\in H_\ell} S_i.
\]

\begin{example}
    Consider a substrate modification network with the following composite reactions
    \[
    S_\emptyset\to S_1,\quad S_1+S_2\to S_3+S_4, \quad S_4\to S_1+ S_3, \quad S_3\to S_2
    \]
    Then the skeleton of the substrate modification network is a hypergraph $G=(V,R)$ where 
    \begin{itemize}
        \item $V=\{S_\emptyset,S_1,S_2,S_3,S_4\}$ and
        \item  $R=\{(\{S_\emptyset\},\{S_1\}), (\{S_1,S_2\},\{S_3,S_4\}), (\{S_4\},\{S_1,S_3\}), (\{S_3\},\{S_2\})\}$.
    \end{itemize}
    
\end{example}

Furthermore, we allow the set of tail and head vertices (and correspondingly $T_\ell$ and $H_\ell$) to be multisets, where there can be repeated instances of their elements. This enables us to represent reactions where the stoichiometric coefficient of a substrate is greater than one. For example, the reaction $S_1\to 2S_2$ can be represented by the hyperedge $(\{S_1\}, \{S_2,S_2\})$.

\subsection{ Hyperedge Current}\label{sec:current_rep}

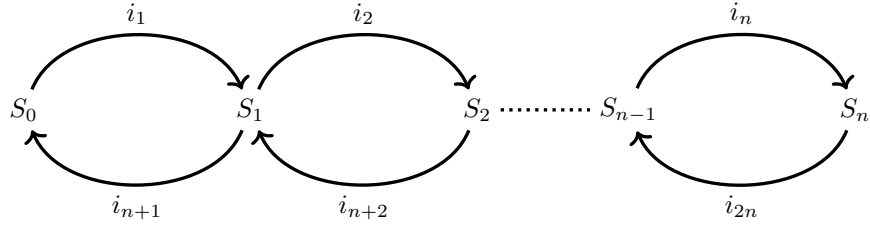
\begin{figure}
    \centering
\begin{tikzpicture}[baseline={(current bounding box.center)},node distance=2cm, on grid]
    \node (S0)  {$S_0$};
    \node (S1) [right=3cm of S0] {$S_1$};
    \node (S2) [right=3cm of S1] {$S_2$};
    \node (Sn-1) [right=2cm of S2] {$S_{n-1}$};
    \node (Sn) [right=3cm of Sn-1] {$S_n$};
    \draw[->,line width=1.25] (S0) .. controls +(0.5,1.23) and +(-0.5,1.23) .. node[above] {$i_1$} (S1);
    \draw[->,line width=1.25] (S1) .. controls +(-0.5,-1.23) and +(0.5,-1.23) .. node[below] {$i_{n+1}$} (S0);
    \draw[->,line width=1.25] (S1) .. controls +(0.5,1.23) and +(-0.5,1.23) .. node[above] {$i_2$} (S2);
    \draw[->,line width=1.25] (S2) .. controls +(-0.5,-1.23) and +(0.5,-1.23) .. node[below] {$i_{n+2}$} (S1);
    \draw[-,dotted,line width=1.25] (S2) .. controls +(1,0) and +(-1,0) .. (Sn-1);
    \draw[->,line width=1.25] (Sn-1) .. controls +(0.5,1.23) and +(-0.5,1.23) .. node[above] {$i_n$} (Sn);
    \draw[->,line width=1.25] (Sn) .. controls +(-0.5,-1.23) and +(0.5,-1.23) .. node[below] {$i_{2n}$} (Sn-1);
\end{tikzpicture}
    \caption{Skeleton of a futile cycle with current assignment.}
    \label{fig:futilecycle_skeleton}
\end{figure}

\begin{figure}
    \centering
    \begin{tikzpicture}[baseline={(current bounding box.center)},node distance=2cm, on grid]
    \node (S0)  {$S_0$};
    \node (S1) [right=3cm of S0] {$S_1$};
    \node (S2) [right=3cm of S1] {$S_2$};
    \node (S3) [right=3cm of S2] {$S_3$};

    \draw[->,line width=1.25] (S0) .. controls +(0.5,1.23) and +(-0.5,1.23) ..  node[above] {$i_1$} (S1);
    \draw[->,line width=1.25] (S1) .. controls +(0.5,1.23) and +(-0.5,1.23) .. node[above] {$i_2$} (S2);
    \draw[->,line width=1.25] (S2) .. controls +(0.5,1.23) and +(-0.5,1.23) .. node[above] {$i_3$} (S3);
    \draw[->,line width=1.25] (S3) .. controls +(-0.5,-1.23) and +(0.5,-1.23) .. node[below] {$i_4$} (S1);
    \draw[->,line width=1.25] (S2) .. controls +(-0.5,-1.23) and +(0.5,-1.23) .. node[below] {$i_5$} (S0);
    \end{tikzpicture}
    \caption{Skeleton of a variant of a futile cycle where phosphates are attached singly but detached in pairs. Current assignment is also shown.}
    \label{fig:jump_return}
\end{figure}
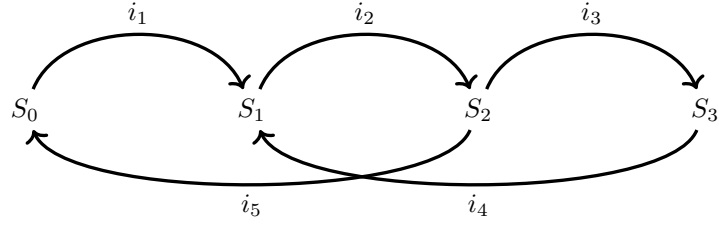

\begin{definition}
Consider a substrate modification network and its skeleton $G=(V,R)$. In a \textit{current assignment}, we associate each edge $\ell\in R$ with a non-negative quantity $I_\ell$ (called \textit{current}) which satisfy the following balance property. For each vertex $S_i$  (excluding $S_\emptyset$) we must have
\begin{equation}\label{eq:vertexbalance}
\sum_{\ell:i\in T_\ell} I_\ell = \sum_{\ell: i\in H_\ell } I_\ell
\end{equation}
In addition, a \textit{positive current assignment} is a current assignment in which $I_\ell>0$ for all $\ell\in R$.   
\end{definition}

\begin{definition}
We say that two hyperedges $(\ell,m)$ are \textit{current-matching} (or equivalently they form a current-matching pair) if there is a positive current assignment and for any choice of positive currents, we have $I_\ell = I_{m}$.  
\end{definition}

\begin{example}
Consider the n-step futile cycle in Figure \ref{fig:futilecycle_skeleton}, where we enumerate all hyperedges and their currents. In any current assignment, we must have
\begin{align*}
    i_1&=i_{n+1} \mbox{ (balance at $S_0$)},\\
    i_1+i_{n+2}&=i_{2}+i_{n+1} \mbox{ (balance at $S_1$)},\\
    &\dots\\
    i_k+i_{n+k+1}&=i_{k+1}+i_{n+k} \mbox{ (balance at $S_k$)},\\
    &\dots\\
   i_n&=i_{2n} \mbox{ (balance at $S_n$)}.
\end{align*}
Thus, it is clear that $(1,n+1)$ is a current-matching pair, and so is $(n,2n)$. We can ``propagate" either of these facts to obtain $i_{k}=i_{n+k}$ for all $k\in[1,n]$. Thus $(k,n+k)$ is a current-matching pair for each $k\in[1,n]$. 
\end{example}

\begin{example}
    Consider the substrate modification networks whose skeletons are given by Figure \ref{fig:cyclic_transfer}. 
    In any current assignment, we must have
    \begin{align*}
       i_1&=i_2 \mbox{ (balance at $X$ or $X_p$)},\\
        i_2&=i_3 \mbox{ (balance at $Y$ or $Y_p$)}, \\
          i_3&=i_1 \mbox{ (balance at $Z$ or $Z_p$)}.
    \end{align*}
    Thus $i_1=i_2=i_3$, i.e. all hyperedges have the same current, and any pair of hyperedges is a current-matching pair.
\end{example}

\subsection{Reaction networks} \label{sec:reactionnetwork}
Recall that each composite reaction, for example $S \to S^*$, can have many different detailed models. For example, a common detailed model of the transformation reaction $S \to S^*$ uses the Henri-Michaelis-Menten mechanism 
\[
S+E \xrightleftarrows{}{} C \longrightarrow S^*+E.
\]
A detailed model can be considered a directed graph whose vertices are linear combination of the species. We emphasize here that this type of directed graph is very different from the skeleton hypergraph that we have been discussing. In order to formally define the former and the dynamical system arising from a detailed model, in this section we present some basic setup and definitions from reaction network theory.

A {\em reaction network} $\GG$ is a directed graph whose vertices are non-negative linear combinations of {\em species} $X_1,X_2, \ldots, X_n$. In the context of this paper, the species are the substrates, enzymes, and compounds which we will describe in Section \ref{sec:detailedmodel}. 

In reaction network literature, we often refer to each vertex as a {\em complex}, and we denote a complex by $y= y_{1}X_1 + y_{2} X_2 + \dots + y_{n} X_n $  or by $y=( y_{1}, y_{2}, \dots, y_{n})$ 	(where $y_{i}\in\Z_{\geq 0}$).

Edges of $\GG$ represent the possible changes in the abundances of the species, and are referred to as {\em reactions}. 
The vector $y'-y$ is the {\em reaction vector} associated to the reaction $y \to y'$. Additionally, in this reaction,  $y$ is called the \textit{reactant complex}, and $y'$ is called the \textit{product complex}. If there is also a reaction from $y'$ to $y$, we write $y \rightleftarrows y'$ and say that they are a pair of {\em reversible reactions}.

\begin{example} \label{ex:futile}
Consider a simple substrate modification network $S\rightleftarrows S^*$, with enzymes $E_1, E_2$, and Henri-Michaelis-Menten detailed model (for both composite reactions). The resulting reaction network is
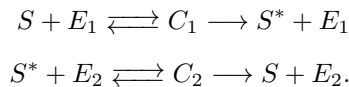
\begin{align*}
S+E_1 \xrightleftarrows{}{} C_1 \longrightarrow S^*+E_1 \\
S^*+E_2 \xrightleftarrows{}{} C_2 \longrightarrow S+E_2.
\end{align*}
Here $\GG$ has 6 species $\{S,S^*,E_1,E_2,C_1,C_2\}$. We call the species $C_1, C_2$ compounds of the two corresponding composite reactions. The reaction network has 6 complexes $\{S+E_1,C_1,S^*+E_1,S^*+E_2,C_2,S+E_2\}$; and 6 reactions (one reaction for each arrow).
\end{example}
Under the assumption of mass-action kinetics, each reaction network $\GG$ defines a  parametrized family of systems of ordinary differential equations (ODEs), as follows. 
First, let $\mathcal{R}$ be the set of all reactions and $r$ be the number of reactions. For each reaction $y\to y'$, we assign a positive {\em rate constant} $\kappa_{y\to y'}\in \mathbb{R}_{>0}$.

Then the \textit{mass-action system}, denoted by $(\GG,\kappa)$ is the dynamical system arising from the following ODEs:
\begin{equation}\label{eq:mass_action_ODE}
		\frac{dx}{dt} ~=~ \sum_{y\to y'\in\mathcal{R}}  \kappa_{y\to y'} x^{y} (y'-y)~=:~ f_{\kappa}(x)~,
\end{equation}
where  $x(t)=(x_1(t),\dots,x_n(t))$ denotes the concentration of the species at time $t$ and $x^{y} := \prod_{i=1}^nx_i^{y_{i}}$. The right-hand side of the ODEs~\eqref{eq:mass_action_ODE} consists of polynomials 
$f_{\kappa,i}$, for $i=1,2,\dots,n$ (where $n$ is the number of species). For simplicity, we often write $f_i$ instead of $f_{\kappa,i}$. 
The ODEs \eqref{eq:mass_action_ODE} can also be written in matrix form
\begin{equation}
    \frac{dx}{dt} = N \cdot v(x),
\end{equation}
where $N$, {\em the stoichiometric matrix}, is the matrix whose columns are all reaction vectors of $\GG$ and  $v_{y\to y'}(x)=\kappa_{y\to y'} x^{y}$. A {\em conservation law matrix} of $\GG$, denoted by $W$, is any row-reduced matrix whose rows form a basis of $\text{im}(N)^\perp$. The {\em conservation laws} of $\GG$ are given by $W x = c$, where $c:=W x(0)$ is the {\em total-constant vector}.

We denote by $S:=\text{im}(N)$, the \textit{stoichiometric subspace} of $\GG$.  Observe that the vector field of the mass-action ODEs~\eqref{eq:mass_action_ODE} lies in $S$ (more precisely, the vector of ODE right-hand sides is always in $S$).  Hence,  a forward-time solution $\{x(t) \mid t \ge 0\}$ of~\eqref{eq:mass_action_ODE}, with initial condition $x(0)  \in \mathbb{R}_{>0}^n$, remains in the following \textit{stoichiometric compatibility class}~\cite{feinberg2019foundations}:$$P_{x(0)} ~:=~ (x(0)+S)\cap \mathbb{R}_{\geq 0}^n~.$$

A {\em steady state} of a mass-action system is a nonnegative concentration vector  $x^*\in \R_{\geq 0}^n$ at which the right-hand side of the ODEs \eqref{eq:mass_action_ODE} vanishes: $f_{\kappa}(x^*)=0$. 

\begin{definition}
$\GG$ is a {\em consistent reaction network} if there exist $\beta_{y \to y'} >0$ such that $\ds \sum_{y \to y' \in \RR} \beta_{y \to y'} (y' - y) = 0$.
\end{definition}

The following theorem  in \cite{joshi2024bifunctional} links consistency and existence of a positive steady state.

\begin{theorem}
The following are equivalent
\been
\item $\GG$ is a consistent reaction network.
\item There is a choice of positive rate constants $\kappa$ such that the mass-action system $(\GG,\kappa)$ has a positive steady state.
\enen
\end{theorem}

In the context of reaction networks, absolute concentration robustness (ACR) can be formally defined at the level of systems and also networks, the latter is a significantly stronger property. 

\begin{definition}[Concentration robustness] \label{def:acr}
Let $X_i$ be a species of a reaction network $\GG$.
\begin{enumerate}
\item For a fixed vector of positive rate constants $\kappa \in \mathbb{R}^r_{>0}$, 
the mass-action system $(\GG,\kappa)$ has {\em absolute concentration robustness} (ACR) in $X_i$ if $(\GG,\kappa)$ has a positive steady state and the value of $x_i$ (the concentration of $X_i$) is the same in every positive steady state. This value is the \textit{ACR value} of $X_i$. 
\item The reaction network $\GG$ has {\em absolute concentration robustness}  in species $X_i$ if $\GG$ is consistent and  furthermore, for any $\kappa'$ such that the mass-action system $(\GG,\kappa')$ has a positive steady state, $(\GG,\kappa')$ must have ACR in $X_i$. 
\item  In a similar manner, we say the mass-action system $(\GG,\kappa)$ has {\em concentration robustness} in a complex $y=\sum_{i=1}^n a_i X_i, \quad a_i \in \Z_{\ge 0}$, if $(\GG,\kappa)$ has a positive steady state and the product
$
x^y:=\prod_{i=1}^n  x_i^{a_i}
$
takes the same value in every positive steady state. The reaction network $\GG$ has {\em concentration robustness}  in complex $y$ if $\GG$ is consistent and furthermore, for any $\kappa'$ such that the mass-action system $(\GG,\kappa')$ has a positive steady state, $(\GG,\kappa')$ has concentration robustness in $y$. 
\end{enumerate}
\end{definition}

Note that the ACR value is independent of the positive steady state of $(\GG,\kappa)$, but this value does depend on $\kappa$.
A natural interpretation of ACR is that a particular steady state coordinate (corresponding the ACR species $X_i$) is independent of initial concentrations.

Finally, we provide some relevant definitions and results from reaction network theory, which we use in later sections.

\begin{definition}
    A reaction network $\GG$ is \textit{weakly reversible} if every connected component is strongly connected.
\end{definition}

\begin{definition}
    The \textit{deficiency} of a reaction network $\GG$ is defined as
    \[
    \delta = \#\text{complexes} - \#\text{connected components} - \text{dim}(S).
    \]
\end{definition}

\begin{theorem}[The deficiency zero theorem \cite{horn1972necessary}]\label{thm:deficiencyzero}
    If $\GG$ is weakly reversible and has a deficiency of 0, then for any choice of rate constant $\kappa$, the mass-action system $(\GG,\kappa)$ has a unique positive steady state in any stoichiometric compatibility class.
\end{theorem}

\subsection{Detailed models}\label{sec:detailedmodel}

Recall from Section \ref{sec:detailed_model} that a commonly used detailed model for $S \to S^*$ is 
\begin{align*} 
S + E~  \xrightleftarrows{}{} ~C \longrightarrow S^* + E,
\end{align*}
where $E$ is an {\em enzyme} that facilitates the reaction while $C$ is an intermediate {\em compound}. 
A detailed model may have more than one intermediate compound, going through a series of steps before the enzyme and the product are released. 
The simplest instance of this has two compounds and two irreversible reactions leading to the final product release
\begin{align*} 
S + E~  \xrightleftarrows{}{} ~C_1 \longrightarrow ~C_2 \longrightarrow S^* + E. 
\end{align*}

The nature of the detailed model is relevant when describing the dynamics, even if one is only interested in the dynamics of the substrates involved. 
In \cite{berezhkovskii2017dependence}, different detailed models that each assume two types of intermediate compound ($C_1$ and $C_2$), as shown in  \eqref{eq:rxnsteps_altmix}, make different predictions for the rates of product formation.

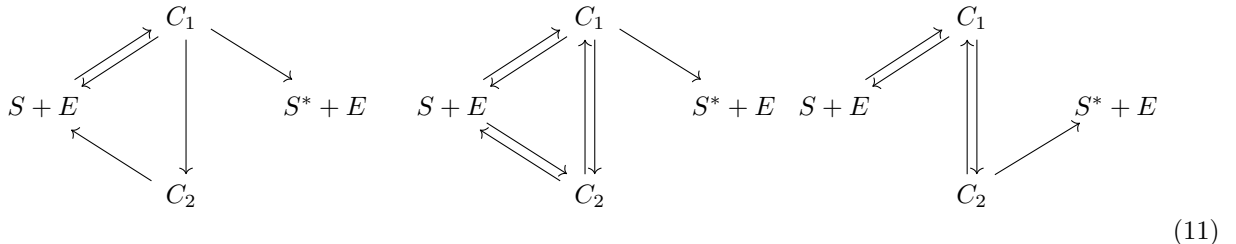
\begin{equation} \label{eq:rxnsteps_altmix}
\begin{aligned}
&\begin{tikzcd}
&  C_1 \arrow[ld,xshift=0.4ex] \arrow[dd,xshift=0.425ex] \arrow[rd,xshift=0.2ex,yshift=0.6ex] \\
S+E \arrow[ru,xshift=-0.2ex,yshift=0.6ex]  &&  S^*+E   \\ 
&  C_2 \arrow[lu,xshift=-0.2ex,yshift=-0.2ex] 
\end{tikzcd} 
& \begin{tikzcd}
&  C_1 \arrow[ld,xshift=0.4ex] \arrow[dd,xshift=0.425ex] \arrow[rd,xshift=0.2ex,yshift=0.6ex] \\
S+E \arrow[ru,xshift=-0.2ex,yshift=0.6ex] \arrow[rd,xshift=0.4ex,yshift=0.4ex] &&  S^*+E   \\ 
&  C_2 \arrow[lu,xshift=-0.2ex,yshift=-0.2ex] \arrow[uu,xshift=-0.425ex]  
\end{tikzcd} 
&  \begin{tikzcd}
&  C_1 \arrow[ld,xshift=0.4ex] \arrow[dd,xshift=0.425ex]  \\
S+E \arrow[ru,xshift=-0.2ex,yshift=0.6ex]  &&  S^*+E   \\ 
&  C_2  \arrow[uu,xshift=-0.425ex] \arrow[ru,xshift=-0.4ex,yshift=0.4ex] 
\end{tikzcd}
\end{aligned}
\end{equation}

Detailed reaction mechanisms involving more than two intermediate compounds are studied in \cite{xu2012realistic}. 
The possibilities grow combinatorially as the number of conformations of the intermediary enzyme-substrate compounds under consideration are increased. 
Even the simplest chain structure can result in many different detailed models, as the following representative examples on just 3 intermediate compounds show. 
\begin{align} \label{eq:rxnsteps11}
S + E~  \xrightleftarrows{}{} ~C_1 \longrightarrow ~ C_2 \longrightarrow  ~C_3 \longrightarrow S^* + E, \\ 
S + E~  \xrightleftarrows{}{} ~C_1 \longrightarrow C_2 \xrightleftarrows{}{} C_3 \longrightarrow  S^* + E. 
\end{align}

An irreversible substrate modification with two intermediate conformations that nevertheless allows for reversible enzyme binding with both the substrate and the product \cite{xu2012realistic,prabakaran2012post,walsh2006posttranslational} is the following
\begin{align} \label{eq:rxnsteps_alt4}
S + E~  \xrightleftarrows{}{} ~&C \longrightarrow  C' ~  \xrightleftarrows{}{} ~ S^* + E. 
\end{align}
A substrate-enzyme compound may not necessarily be an intermediary in the process of product formation. 
An example of this is seen in the detailed reaction mechanism for the EnvZ-OmpR model with ADP as cofactor presented by \cite{shinar2010structural}.
The relevant subnetwork is shown below, see also Section \ref{sec:skeleton} for the complete network:  
\begin{align} \label{eq:rxnsteps_alt5}
C' \xrightleftarrows{}{} ~& S + E~  \xrightleftarrows{}{} ~C \longrightarrow   ~ S^* + E. 
\end{align}

When considering substrate modification, even for a simple composite reaction $S\to S^*$, the possibilities of detailed models are unlimited. 
We wish to consider a sufficiently general mathematical framework that has the flexibility to accommodate all the detailed models of substrate modification considered above, among many others.

We now proceed to give precise description on the detailed models we consider in this paper. 
\begin{assumption}\label{assum:blocks}
An allowed detailed model for each composite reaction $\ell$: $\sum_{i\in T_\ell} S_i\to \sum_{i\in H_\ell} S_i$ with the enzyme $E$ ($E$ denotes the null enzyme when there is no enzyme) must satisfy the following: 
\begin{enumerate}
    \item There is a {\em product block}, which is a strongly connected component with one complex $\sum_{i\in H_\ell} S_i+E$ and the rest of the complexes (if any) are single species complexes.  The single species complexes in the product block will be called the {\em product compounds}. 
    \item The  induced subgraph obtained from the subset of complexes that are {\em outside} the product block will be called the {\em reactant block}. The reactant block is a (weakly) connected subgraph that contains one complex $\sum_{i\in T_\ell} S_i+E$ and the rest are single species complexes, which we call {\em reactant compounds}.

    \item There is a single edge from a complex in the reactant block to a complex in the product block. The former complex will be called the {\em terminal complex of the reactant block}.  

    \item Every edge in the reactant block is either part of a directed cycle or is part of a directed path of edges from $\sum_{i\in T_\ell} S_i+E$ to the terminal complex of the reactant block. (One way to satisfy this property is if the reactant block is strongly connected.) 

\end{enumerate}
\end{assumption}

Assumption \ref{assum:blocks} (3) can be relaxed; if there are more than one edges from the reactant block to the product block, it may be possible to split the composite reaction into multiple hyperedges between the same tail and head vertices. Assumption \ref{assum:blocks} (4) is  necessary (but may not be sufficient) for the existence of a positive steady state, i.e. if it is not satisfied then there cannot be a positive steady state.

It is a simple matter to check that every detailed model above satisfies these assumptions. 
It is instructive to clearly identify the reactant block and the product block, which we depict as being enclosed in blue box and a red box, respectively.

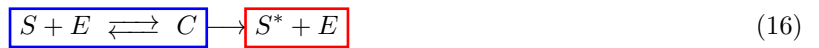
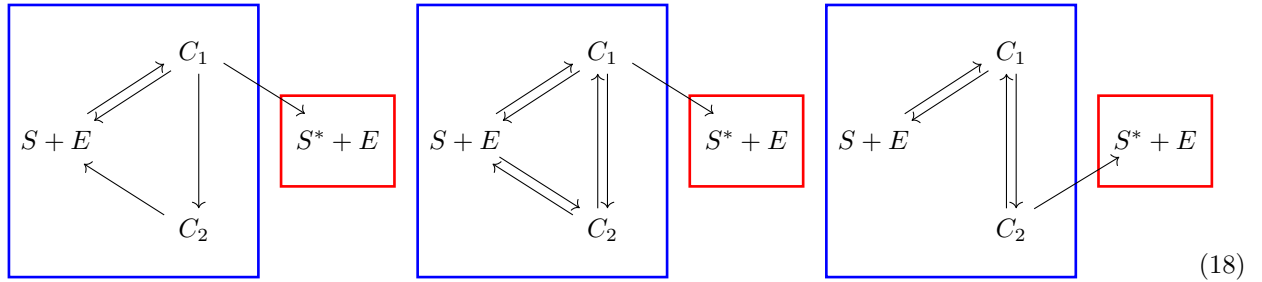
\def\tikzmark#1{\tikz[remember picture,overlay]\node[inner ysep=1pt,anchor=base](#1){\strut};}

\begin{equation}
\label{eq:rxnsteps_block}
\tikzmark{A} S + E~  \xrightleftarrows{}{} ~C \tikzmark{B} \longrightarrow \tikzmark{C} S^* + E \tikzmark{D}
\tikz[remember picture,overlay]\draw[draw=blue,inner xsep=3mm, inner ysep=2mm,line width=1](A.south west)rectangle(B.north east);
\tikz[remember picture,overlay]\draw[draw=red,inner xsep=3mm, inner ysep=2mm,line width=1](C.south west)rectangle(D.north east);
\end{equation} 

\begin{align} \label{eq:rxnsteps6_block}
\tikzmark{A} S + E~  \xrightleftarrows{}{} ~C_1 \longrightarrow ~C_2 \tikzmark{B} \longrightarrow \tikzmark{C} S^* + E \tikzmark{D} 
\tikz[remember picture,overlay]\draw[draw=blue,inner xsep=3mm, inner ysep=2mm,line width=1](A.south west)rectangle(B.north east);
\tikz[remember picture,overlay]\draw[draw=red,inner xsep=3mm, inner ysep=2mm,line width=1](C.south west)rectangle(D.north east);
\end{align}

\begin{equation} \label{eq:rxnsteps_altmix_block}
\begin{aligned}
& 
\tikz[overlay]{
    \draw[draw=blue,inner xsep=3mm, inner ysep=2mm,line width=1] (0,-1.8) rectangle (3.3,1.8);
    \draw[draw=red,inner xsep=3mm, inner ysep=2mm,line width=1] (3.6,-.6) rectangle (5.1,.6);
}
\begin{tikzcd}
&  C_1 \arrow[ld,xshift=0.4ex] \arrow[dd,xshift=0.425ex] \arrow[rd,xshift=0.2ex,yshift=0.6ex] \\
S+E \arrow[ru,xshift=-0.2ex,yshift=0.6ex]  &&  S^*+E   \\ 
&  C_2 \arrow[lu,xshift=-0.2ex,yshift=-0.2ex] 
\end{tikzcd} 
& 
\tikz[overlay]{
    \draw[draw=blue,inner xsep=3mm, inner ysep=2mm,line width=1] (0,-1.8) rectangle (3.3,1.8);
    \draw[draw=red,inner xsep=3mm, inner ysep=2mm,line width=1] (3.6,-.6) rectangle (5.1,.6);
}
\begin{tikzcd}
&  C_1 \arrow[ld,xshift=0.4ex] \arrow[dd,xshift=0.425ex] \arrow[rd,xshift=0.2ex,yshift=0.6ex] \\
S+E \arrow[ru,xshift=-0.2ex,yshift=0.6ex] \arrow[rd,xshift=0.4ex,yshift=0.4ex] &&  S^*+E   \\ 
&  C_2 \arrow[lu,xshift=-0.2ex,yshift=-0.2ex] \arrow[uu,xshift=-0.425ex]  
\end{tikzcd} 
& \quad
\tikz[overlay]{
    \draw[draw=blue,inner xsep=3mm, inner ysep=2mm,line width=1] (0,-1.8) rectangle (3.3,1.8);
    \draw[draw=red,inner xsep=3mm, inner ysep=2mm,line width=1] (3.6,-.6) rectangle (5.1,.6);
}
\begin{tikzcd}
&  C_1 \arrow[ld,xshift=0.4ex] \arrow[dd,xshift=0.425ex]  \\
S+E \arrow[ru,xshift=-0.2ex,yshift=0.6ex]  &&  S^*+E   \\ 
&  C_2  \arrow[uu,xshift=-0.425ex] \arrow[ru,xshift=-0.4ex,yshift=0.4ex] 
\end{tikzcd}
\end{aligned}
\end{equation}

\begin{align} \label{eq:rxnsteps11_block}
\tikzmark{A} S + E~  \xrightleftarrows{}{} ~C_1 \longrightarrow ~ C_2 \longrightarrow  ~C_3 \tikzmark{B} \longrightarrow \tikzmark{C} S^* + E \tikzmark{D}, \\ 
\tikz[remember picture,overlay]\draw[draw=blue,inner xsep=3mm, inner ysep=2mm,line width=1](A.south west)rectangle(B.north east);
\tikz[remember picture,overlay]\draw[draw=red,inner xsep=3mm, inner ysep=2mm,line width=1](C.south west)rectangle(D.north east);
\tikzmark{A} S + E~  \xrightleftarrows{}{} ~C_1 \longrightarrow C_2 \xrightleftarrows{}{} C_3 \tikzmark{B} \longrightarrow  \tikzmark{C} S^* + E \tikzmark{D}.
\tikz[remember picture,overlay]\draw[draw=blue,inner xsep=3mm, inner ysep=2mm,line width=1](A.south west)rectangle(B.north east);
\tikz[remember picture,overlay]\draw[draw=red,inner xsep=3mm, inner ysep=2mm,line width=1](C.south west)rectangle(D.north east);
\end{align}

\begin{align} \label{eq:rxnsteps_alt4_block}
\tikzmark{A} S + E~  \xrightleftarrows{}{} ~&C \tikzmark{B}  \longrightarrow  \tikzmark{C} C' ~  \xrightleftarrows{}{} ~ S^* + E \tikzmark{D} 
\tikz[remember picture,overlay]\draw[draw=blue,inner xsep=3mm, inner ysep=2mm,line width=1](A.south west)rectangle(B.north east);
\tikz[remember picture,overlay]\draw[draw=red,inner xsep=3mm, inner ysep=2mm,line width=1](C.south west)rectangle(D.north east);
\end{align}
 
\begin{align} \label{eq:rxnsteps_alt5_block}
\tikzmark{A} C' \xrightleftarrows{}{} ~& S + E~  \xrightleftarrows{}{} ~C \tikzmark{B} \longrightarrow   ~ \tikzmark{C}  S^* + E \tikzmark{D}  
\tikz[remember picture,overlay]\draw[draw=blue,inner xsep=3mm, inner ysep=2mm,line width=1](A.south west)rectangle(B.north east);
\tikz[remember picture,overlay]\draw[draw=red,inner xsep=3mm, inner ysep=2mm,line width=1](C.south west)rectangle(D.north east);
\end{align}

The compounds $C_1$ and $C_2$ are in the reactant block for all three detailed models in \eqref{eq:rxnsteps_altmix_block} since the induced graph on the complexes $S+E$, $C_1$ and $C_2$ is a strongly connected component. 
The same is true for \eqref{eq:rxnsteps6_block} even though the induced graph on these three complexes is not a strongly connected component. 
The reason is that the edge from $C_1$ to $C_2$ is part of the directed path from $S+E$ to the terminal vertex $C_2$ of the reactant block, thus satisfying the third requirement. 
An interesting distinction must be made for \eqref{eq:rxnsteps_alt4_block}, where the compound $C$ (but not $C'$) is in the reactant block. In fact, $C'$ belongs to the product block. 

Finally note that the reversible Michaelis-Menten reaction \eqref{eq:revMM_rep} fails to satisfy the above assumptions because the entire reaction network is one strongly connected component, which does not allow separation into reactant and product blocks.  
\begin{align} \label{eq:revMM_rep}
S + E~  \xrightleftarrows{}{} ~C \xrightleftarrows{}{} ~S^* + E
\end{align}
In fact, due to the symmetry between the roles of $S$ and $S^*$ in such a scheme, the labeling of one substrate as reactant and another as product is arbitrary. 

So far we have only discussed the assumptions on the detailed model for a single edge in the skeleton. 
Finally, we state the assumption on the detailed model required when putting the skeleton edges together. Essentially, we only need that different skeleton edges do not associate with the same compound in the overall detailed model. This is a  reasonable assumption for almost any biochemical model and therefore not restrictive in application. 
\begin{assumption}\label{assumption:no_repeat_compounds}
Let $C_{\{\ell\}}$ denote the set of compounds associated with the hyperedge $\ell$. 
If $\ell \ne m$ then 
$C_{\{\ell\}} \bigcap C_{\{m\}} = \varnothing$.
\end{assumption}

\subsection{Enzymes}\label{sec:enzyme}
Since some composite reactions may use bifunctional enzymes, and multiple composite reactions may have the same enzyme, it is useful to have a flexible notation for the enzyme in each hyperedge. In this section, we will set up such a notation.

First, let the (non-bifunctional) enzymes in the network be $E_1,E_2,\dots$. To represent the composite reactions without an enzyme in a consistent manner, we introduce the ``null enzyme" $E_\emptyset = 0$. Besides these enzymes, we also consider bifunctional enzymes, which are among the compounds $C_1,C_2,\dots$. Next, we introduce a function $\mathcal{E}: R\to \{E_\emptyset,E_1,\dots, C_1,C_2,\dots\}$ which maps each hyperedge $\ell$ to the enzyme acting in the corresponding composite reaction. For the sake of clarity, here we provide some examples to illustrate this notation:
\begin{itemize}
    \item If hyperedge $\ell$ uses the enzyme $E_1$, then we write $\mathcal{E}_\ell = E_1$. Since multiple composite reactions may have the same enzyme, we may have $\EE_\ell=\EE_m$ for $\ell\neq m$ (i.e. the map $\EE$ is not necessarily injective).
    \item If hyperedge $\ell$ does not have an enzyme, then from the convention of null enzyme earlier, we write $\mathcal{E}_\ell=E_\emptyset$.
    \item If hyperedge $\ell$ uses a bifunctional enzyme, which is the compound $C_1$, then we write $\mathcal{E}_\ell = C_1$.
\end{itemize}

\section{Substrate modification networks with Henri-Michaelis-Menten detailed model}\label{sec:HMM}

For this section, to build intuition we assume that the detailed model for each composite reaction has the Henri-Michaelis-Menten form (Figure \ref{eq:rxnsteps_block}). More general detailed models will be considered in Section \ref{sec:general}.

\subsection{Detailed model}

We consider a substrate modification network with $n$ substrates $S_1,S_2,\dots, S_n$ and $d$ (non-bifunctional) enzymes $E_1,\dots,E_d$. Recall that we may use the null substrate $S_\emptyset$ and null enzyme $E_\emptyset$ if necessary.
Let $G=(V,R)$ be the skeleton of the substrate modification network. Each hyperedge $\ell \in R$ corresponds to the composite reaction
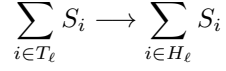
\[
\sum_{i\in T_\ell} S_i \longrightarrow \sum_{i\in H_\ell} S_i 
\]
and its Henri-Michaelis-Menten detailed model is
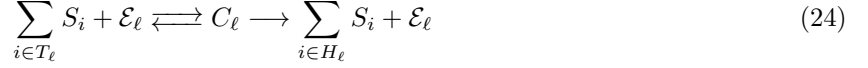
\begin{align}\label{eq:HMMdetailedmodel}
\sum_{i\in T_\ell} S_i + \EE_\ell \xrightleftarrows{}{} C_\ell \longrightarrow \sum_{i\in H_\ell} S_i + \EE_\ell
\end{align}
Here we remind the readers that $\EE$ is a function mapping each hyperedge to its enzyme.

\subsection{Resulting mass-action ODEs}\label{sec:ODEs}

After specifying the detailed model (Henri-Michaelis-Menten) for all hyperedges, we have associated the substrate modification network and its skeleton $G$ with a reaction network $\mathcal{G}$. This section discusses the mass-action ODEs arising from $\mathcal{G}$.

\vspace{.1in}
\noindent \textbf{Without bifunctional enzymes}
\vspace{.1in}

The mass-action ODEs arising from $\GG$ (with $n$ substrates), where there are $d$ enzymes (none of which are bifunctional), are given as follows
\begin{equation}\label{eq:non-bifunctional-HMM}
\begin{aligned}
&\frac{dc_{\ell}}{dt}  = G_{\ell}    &&\text{for}\quad \ell\in R\\
&\frac{de_j}{dt}    = -\left(\sum_{\ell: \mathcal{E}_\ell=E_j}  G_\ell\right) &&\text{for} \quad j\in [1,d]\\
&\frac{ds_i}{dt} = L_i- \left(\sum_{\ell: i\in T_\ell} G_{\ell}\right) &&\text{for}\quad i\in [1,n]  
\end{aligned}
\end{equation}
where
\begin{equation}\label{eq:non-bifunctional-HMM-GL}
\begin{aligned}
& G_{\ell} = k_{\ell}^+ (\prod_{i\in T_\ell} s_i)\varepsilon_\ell - (k_\ell^- +k_\ell)c_\ell, \quad &&\text{for} \quad \ell \in R \\
& L_i =  \sum_{\ell: i\in H_\ell} k_{\ell} c_{\ell} - \sum_{\ell: i\in T_\ell} k_{\ell}c_{\ell} , \quad &&\text{for} \quad i\in[1,n]
\end{aligned}
\end{equation}
Here $\varepsilon$ (corresponding to $\mathcal{E}$) is a function mapping each hyperedge to its enzyme concentration. For example, if $E_1$ is the enzyme in hyperedge $\ell$, i.e. $\mathcal{E}_\ell = E_1$, then  $\varepsilon_\ell=e_1$ and $G_\ell= k_{\ell}^+ (\prod_{i\in T_\ell} s_i)e_1 - (k_\ell^- +k_\ell)c_\ell$.

For the null substrate $S_\emptyset$ and null enzyme $E_\emptyset$, we make the convention that $s_\emptyset=e_\emptyset=1$ at all times, and thus $ds_\emptyset/dt=de_\emptyset/dt=0$.

The expressions $G_\ell$ can be interpreted as the compound production rate, or the enzyme consumption rate on hyperedge $\ell$. 
The expressions $L_i$ will be interpreted later as the algebraic sum of the currents at the substrate $S_i$ in the skeleton.

\vspace{.1in}

\noindent\textbf{With or without bifunctional enzymes}

\vspace{.1in}
 
 In the (possible) presence of bifunctional enzyme(s), where a compound in one edge may act as enzyme in another edge, the mass-action ODEs are given as follows

\begin{equation}\label{eq:bifunctional-hmm}
\begin{aligned}
&\frac{dc_{\ell}}{dt}  = G_{\ell}  + \widetilde{G}_\ell  \quad &&\text{for}\quad \ell\in R\\
&\frac{de_j}{dt}    = -\left(\sum_{\ell: \mathcal{E}_\ell=E_j}  G_\ell\right) \quad &&\text{for} \quad j\in [1,d]\\
&\frac{ds_i}{dt} = L_i- \left(\sum_{\ell: i\in T_\ell} G_{\ell}\right)\quad &&\text{for}\quad i\in [1,n] 
\end{aligned}
\end{equation}
where $G_\ell$ and $L_i$ are defined same as in \eqref{eq:non-bifunctional-HMM-GL} and 

\begin{equation}\label{eq:bifunctional-hmm-Gtilde}
\begin{aligned}
&\widetilde{G}_\ell = -\left(\sum_{m\in R: \mathcal{E}_m=C_\ell}  G_m\right).
\end{aligned}
\end{equation}

The two terms $G_\ell$ and $\widetilde{G}_\ell$ will be referred to as the {\em local part} and the {\em nonlocal part} of the influence on $c_\ell$. 
If $\widetilde{G}_\ell=0$, $C_\ell$ is {\em not} a bifunctional enzyme. 

Additionally, with bifunctional enzymes, some $\varepsilon_\ell$ may be assigned to compound concentrations. For example, if $C_1$ is the enzyme for hyperedge $\ell$, then we have $\varepsilon_\ell=c_1$ and thus $G_\ell= k_{\ell}^+ (\prod_{i\in T_\ell} s_i)c_1 - (k_\ell^- +k_\ell)c_\ell$.

\vspace{.1in}

\subsection{Analysis of the steady state equations}\label{sec:analysis_sseq}

In this section, we discuss the steady states of substrate modification networks with or without a bifunctional enzyme (i.e. from the ODEs \eqref{eq:bifunctional-hmm}). First, we make the following assumption.
\begin{assumption}\label{assumption:no_conservation_compounds}
There is no conservation law which contains only the compounds $\{c_\ell: \ell\in R\}$.
\end{assumption}

\begin{remark}
While Assumption \ref{assumption:no_conservation_compounds} is used in the proofs for our main results, it is not purely technical. It is easy to check that without bifunctional enzymes, the assumption always holds. With bifunctional enzymes and under reasonable biological models, the assumption has ample physical justification. 

To illustrate this, we first present the simplest case when the assumption is violated:
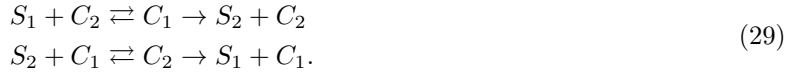
\begin{equation}\label{model:assum1}
\begin{aligned} 
&S_1 + C_2 \rightleftarrows C_1 \to S_2 + C_2\\
&S_2 + C_1 \rightleftarrows C_2 \to S_1 + C_1.
\end{aligned}   
\end{equation}

The conservation law $c_1+c_2=T$ violates the requisite assumption. 
However, note the unphysicality of the model. Since $C_1$ is the bound form of $S_1$ and $C_2$ while $C_2$ is the bound form of $S_2$ and $C_1$, it leads to the bizarre scenario of the compound $C_1$ being a proper component of $C_1$ itself! 

In fact, as we will argue later, in some cases Assumption \ref{assumption:no_conservation_compounds} can be weakened without impacting the theoretical results. 
It will turn out that the theory can even be applied to the network \ref{model:assum1} without any problem.  
\end{remark}

We first show that under Assumption \ref{assumption:no_conservation_compounds}, there is a natural decompostion of the steady state equations.
\begin{lemma}\label{lem:Fij}
The steady state equations for \eqref{eq:bifunctional-hmm} are equivalent to
\begin{alignat*}{2}
&G_\ell=0\quad &&\text{for}\quad \ell\in R\\
&L_i= 0 \quad &&\text{for}\quad i\in[1,n]. 
\end{alignat*}
\end{lemma}
\begin{proof}
By assumption \ref{assumption:no_conservation_compounds}, for $\alpha_\ell \in \R$, if 
$
\frac{d}{dt} \sum_{\ell \in R} \alpha_\ell c_\ell = 0  
$
then $\alpha_\ell = 0$ for all $\ell \in R$.
In other words, the set 
\[
\left\{ \frac{dc_\ell}{dt}: \ell \in R \right\} = \left\{ G_\ell + \widetilde{G}_\ell: \ell \in R \right\} 
\]
is linearly independent. 

Now at steady state, this results in a homogeneous system of $\# R$ linearly independent equations in $\# R$ variables $\{G_\ell : \ell \in R\}$, from which we conclude that $G_\ell = 0$ for all $\ell \in R$. 
Combining this with $ds_i/dt=0$ yields $L_i=0$ for all $i\in[1,n]$.  The other direction is trivial.
\end{proof}

\begin{remark}
It is possible in some cases to relax Assumption \ref{assumption:no_conservation_compounds} and still have the results in Lemma \ref{lem:Fij}. We will illustrate this with the network \eqref{model:assum1}. We have
\[
\frac{dc_1}{dt}=G_1 - G_2, \quad \frac{dc_2}{dt}=G_2-G_1, \quad \frac{ds_1}{dt}=-k_1c_1+k_2c_2 -G_1,\quad \frac{ds_2}{dt}=k_1c_1-k_2c_2 -G_2
\]
and thus $\{ dc_1/dt, dc_2/dt\}$ does not form a linear independent set. However, we can obtain another equation in $G_1,G_2$ by taking certain linear combination of $\{ds_1/dt, ds_2/dt\}$. For this particular network, we can take their sum to obtain
\[
\frac{ds_1}{dt}+\frac{ds_2}{dt}=-G_1-G_2.
\]
Thus $\{dc_1/dt, ds_1/dt+ds_2/dt\}$ forms a linearly independent set and at steady state we still have $G_1=G_2=0$. This allows us to apply Theorem \ref{thm:ACR} and conclude that the network \eqref{model:assum1} has ACR in both $S_1$ and $S_2$. 

In general, it is not trivial to find a linear combination of $\{ds_i/dt: i\in [1,n]\}$ to obtain additional linear combinations of $\{G_\ell:\ell\in R\}$ (especially when there are composite reactions of the types other than $S\to S^*$). As such, we will just focus on Assumption \ref{assumption:no_conservation_compounds} for the rest of the paper.
\end{remark}

The decomposition of the steady state equations in Lemma \ref{lem:Fij} allows us to establish the existence of a positive steady state based on some conditions on the skeleton $G$.

\begin{theorem} \label{thm:consistency}
Consider a substrate modification network where every reaction has the detailed model given in \eqref{eq:HMMdetailedmodel}. Suppose there exists a positive current assignment on the skeleton $G$. Then the associated reaction network $\GG$ is consistent.
\end{theorem}
\begin{proof}
Suppose there exists a positive current assignment $\{I_\ell>0:\ell\in R\}$. We will show that there exists a choice of rate constants $\kappa$ such that the resulting mass-action system admits a positive steady state $x^*$ where the concentration of every species is equal to 1. In particular, we make the following choice of rate constants. 
\begin{itemize}
\item First, we pick $k_{\ell} = I_{\ell}$ for all $\ell\in R$. At steady state $x^*$ where every $c_\ell=1$, we have
\[
L_i =  \sum_{\ell: i\in H_\ell} k_{\ell} c_{\ell} - \sum_{\ell: i\in T_\ell} k_{\ell}c_{\ell}  = \sum_{\ell: i\in H_\ell} I_{\ell}  - \sum_{\ell: i\in T_\ell} I_{\ell} = 0,
\]
where the last equality is due to the balance property of the currents.
\item We pick $k_{\ell}^- = 1$ for all $\ell\in R$. Furthermore, we pick $k_\ell^+=1+I_\ell$ for all $\ell\in R$. With this choice, at steady state $x^*$ where all $s_i=1$, $c_\ell=1$ and $e_j=1$ (thus all $\varepsilon_\ell=1$) we have
\[
G_\ell = k_{\ell}^+ (\prod_{i\in T_\ell} s_i)\varepsilon_\ell - (k_\ell^- +k_\ell)c_\ell = 1+I_\ell - (1+I_\ell)=0.
\]
\end{itemize}
Thus, from Lemma \ref{lem:Fij}, $x^*$ is a positive steady state of the resulting mass-action system.
\end{proof}

Now we turn to the main result of this section on sufficient conditions for ACR. We first provide a formal definition for bifunctional enzyme action.

\begin{definition} \label{def:bifunctional}
A pair of hyperedges $(\ell,m)\in R\times R$ is called a {\em bifunctionally linked pair} if the compound of $\ell$ is the enzyme of $m$ (i.e. $\EE_{m}=C_\ell$). We call $\ell$  the {\em c-edge} and $m$ the {\em e-edge} of the bifunctionally linked pair.
\end{definition}

\begin{theorem}\label{thm:ACR}
Consider a substrate modification network where every reaction has the detailed model given in \eqref{eq:HMMdetailedmodel}. Suppose that the skeleton $G$ contains a pair of hyperedges $(\ell,m)$ that is both  bifunctionally linked and  current-matching. Then the associated reaction network $\GG$ has concentration robustness in the tail of the e-edge $m$. 
Furthermore, if the tail of $m$ contains only one reactant substrate, then the associated reaction network $\GG$ has ACR in that reactant substrate.
\end{theorem}
\begin{proof}
From Lemma \ref{lem:Fij}, at steady state, we have 
\[
L_i =  \sum_{\ell: i\in H_\ell} k_{\ell} c_{\ell} - \sum_{\ell: i\in T_\ell} k_{\ell}c_{\ell}  = 0
\]
for all $i\in [1,n]$. Thus, by setting $I_\ell = k_\ell c_\ell$, the set $\{I_\ell:\ell\in R\}$ forms a positive current assignment on $G$. Since $(\ell,m)$ is  current-matching, their currents must equal for any choice of positive currents. Thus we have
\[
I_\ell = I_m \implies k_\ell c_\ell = k_m c_m \implies c_m = \frac{k_\ell}{k_m} c_\ell
\]
Furthermore, we also have $(\ell,m)$ is a bifunctionally linked pair, which gives $\EE_m=C_\ell$ and $\varepsilon_m=c_\ell$. From Lemma \ref{lem:Fij}, at steady state we must have $G_m=0$, thus
\[
0=G_m = k_{m}^+ (\prod_{i\in T_m} s_i)\varepsilon_m - (k_m^- +k_m)c_m = k_{m}^+ (\prod_{i\in T_m} s_i) c_\ell - (k_m^- +k_m)\frac{k_\ell}{k_m} c_\ell.
\]
Eliminating $c_\ell$ and moving terms, we have
\[
\prod_{i\in T_m} s_i = \frac{k_\ell}{\frac{k_mk^+_m}{k_m^-+k_m}} =: \frac{k_\ell}{k^*_m}
\]
at any steady state. Thus $\GG$ has concentration robustness in the tail of $m$. Furthermore, if there is only one $i\in T_m$, then we must have
\[
s_i = \frac{k_\ell}{k^*_m}
\]
and thus $\GG$ has ACR in substrate $S_i$, the only reactant substrate of the e-edge $m$.

\end{proof}

\begin{remark} \label{remark:non-mass-action}
It is easy to see that Theorem \ref{thm:ACR} can be extended beyond mass-action kinetics, as long as the reaction rates are still linear in the enzyme and compound concentrations. In such a case, we replace the current $G_\ell$ by a more general form:
\[
G_\ell = k_\ell^+ f_\ell(\{s_i:i\in T_\ell\}) - (k_\ell^-+k_\ell)c_\ell.
\]
Following Theorem \ref{thm:ACR}, if a pair of hyperedges $(\ell,m)$ is both bifunctionally linked and  current-matching, then $f_m(\{s_i:i\in T_m\})$ has the same value at any positive steady state. In particular, if $T_m$ only contains a single substrate $S_i$ and $f_m$ is monotone, then $S_i$ is an ACR species (when a positive steady state exists).
\end{remark}

\section{Substrate modification networks with general detailed models}\label{sec:general}

In this section, we show that the techniques and results in Section \ref{sec:HMM} can be extended to substrate modification networks whose composite reactions may have more general detailed models as described in Section \ref{sec:detailedmodel}.

\subsection{Detailed model}

We start by considering a substrate modification network $\mathcal{G}$ with $n$ substrates $S_1,S_2,\dots, S_n$ and $d$ enzymes $E_1,\dots,E_d$. 
As before, some reactions may involve the null substrate $S_\emptyset$ or the null enzyme $E_\emptyset$. 
Let $G=(V,R)$ be the skeleton of the substrate modification network $\GG$. Each hyperedge $\ell \in R$ corresponds to the composite reaction
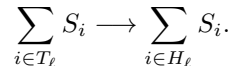
\[
\sum_{i\in T_\ell} S_i \longrightarrow \sum_{i\in H_\ell} S_i.
\]

Assume that each composite reaction has a detailed model satisfying Assumptions \ref{assum:blocks} in Section \ref{sec:detailedmodel}.
By Assumption \ref{assumption:no_repeat_compounds}, the set of indices on all compounds in the overall detailed model are partitioned into $2 \left(\# R\right)$ sets, the reactant and product compounds of each hyperedge. 
Denote by $RC_\ell$ the set of indices on the reactant compounds associated with the hyperedge $\ell$, and therefore the reactant compounds associated with the hyperedge $\ell$ are $\{C_{\ell p}:p\in RC_\ell\}$. 
Similarly, the set of indices on the product compounds for hyperedge $\ell$ are $PC_\ell$ and so the product compounds are $\{C_{\ell p}: p\in PC_\ell\}$. 
Without loss of generality, we denote the terminal complex of the reactant block by $C_{\ell 1}$  (recall from Assumptions \ref{assum:blocks} that a terminal complex always exists).

Lastly, we again use the notation $\EE_\ell$, which indicates the (non-bifunctional or bifunctional)  enzyme for hyperedge $\ell$.

\subsection{Resulting mass-action ODEs}
After specifying the detailed model for all hyperedges, we have associated the substrate modification network and its skeleton $G$ with a reaction network $\mathcal{G}$. This section discusses the mass-action ODEs arising from $\mathcal{G}$. 

The mass action ODEs are given in \eqref{eq:bifunctional-hmm-general} -- it will be convenient to derive them by going from the special case of a single composite reaction (single hyperedge) \eqref{eq:singleedge}, to the case where there is no bifunctional enzyme \eqref{eq:nonbifunctional-hmm-general}, finally to the fully general case \eqref{eq:bifunctional-hmm-general}. 

\vspace{.1in}
\noindent\textbf{Without bifunctional enzymes}
\vspace{.1in}

First we consider the ODEs arising from the detailed model of a single composite reaction. 

\begin{lemma}
    Consider a single composite reaction: $S\to S^*$ with enzyme $E$ where the detailed model satisfies Assumption \ref{assum:blocks}.  Denote the reactant compounds by $\{C_{p}: p\in RC\}$ and the product compounds by $\{C_{p}: p\in PC\}$, where $RC$ and $PC$ are the set of indices of the reactant compounds and product compounds. Let $C_1$ be the terminal complex of the reactant block and $k_1$ be the rate constants for the reaction from $C_1$ to the product block. 
    Then we have the identities
\begin{equation}\label{eq:singleedge}
\begin{aligned}
        &\frac{ds}{dt}   &&=  -k_1 c_1 - \left(\sum_{p\in RC} \frac{dc_p}{dt}\right),\\
        &\frac{ds^*}{dt} &&= k_1c_1   - \left(\sum_{p\in PC} \frac{dc_p}{dt}\right), \quad \mbox{ and }\\
        &\frac{de}{dt} &&= -\left(\sum_{p\in RC\cup PC} \frac{dc_p}{dt}\right).
\end{aligned}
\end{equation}
\end{lemma}
\begin{proof}
    We take the sum of $ds/dt$ and  $dc_p/dt$ for all $p\in RC$. Notice that the contribution of each edge in the reactant block to the sum is zero since the species in all complexes are distinct. In addition, there is a single edge from $C_1$ out of the reactant block with rate constant $k_1$, which results in the first identity. 
    For the second identity, we take the sum of $ds^*/dt$ and $dc_p/dt$ and notice that there is a single edge from $C_1$ into the product block. 
    Finally, since $E$ appears in both the reactant and product blocks, we get the third identity. 
\end{proof}
The argument extends easily to other types of composite reactions, possibly involving multiple substrates on the reactant side or on the product side.

Now we can assemble the mass-action ODEs arising from the entire reaction network $\GG$ where there is no bifunctional enzyme. 
Let $G_{\ell p}$ denote the ODE for each compound $c_{\ell p}$ where $\ell\in R, p\in RC_\ell\cup PC_\ell$. The mass-action ODEs of $\GG$ is given as follows

\begin{equation}\label{eq:nonbifunctional-hmm-general}
\begin{aligned}
&\frac{dc_{\ell p}}{dt}  = G_{\ell p}    \quad &&\text{for}\quad \ell\in R, p\in RC_\ell \cup PC_\ell, \\
&\frac{de_j}{dt}    = -\left(\sum_{\ell: \mathcal{E}_\ell=E_j}  \sum_{p\in RC_\ell\cup PC_\ell}G_{\ell p}\right) \quad &&\text{for} \quad j\in [1,d], \\
&\frac{ds_i}{dt} = L_i- \left(\sum_{\ell: i\in T_\ell} \sum_{p\in RC_\ell}G_{\ell p} +  \sum_{\ell: i\in H_\ell} \sum_{p\in PC_\ell}G_{\ell p}\right)\quad &&\text{for}\quad i\in [1,n],\\
\mbox{where} \\
&L_i =  \sum_{\ell: i\in H_\ell} k_{\ell} c_{\ell 1} - \sum_{\ell: i\in T_\ell} k_{\ell}c_{\ell 1} , \quad &&\text{for} \quad i\in[1,n].
\end{aligned}
\end{equation}

\vspace{.1in}
\noindent\textbf{With bifunctional enzymes}
\vspace{.1in}

In the presence of bifunctional enzyme(s),  the mass-action ODEs are given as follows

\begin{equation}\label{eq:bifunctional-hmm-general}
\begin{aligned}
&\frac{dc_{\ell p}}{dt}  = G_{\ell p}  + \widetilde{G}_{\ell p}  \quad &&\text{for}\quad \ell\in R, p\in RC_\ell\cup PC_\ell, \\
&\frac{de_j}{dt}    = -\left(\sum_{\ell: \mathcal{E}_\ell=E_j}  \sum_{p\in RC_\ell\cup PC_\ell}G_{\ell p}\right) \quad &&\text{for} \quad j\in [1,d], \mbox{ and }\\
&\frac{ds_i}{dt} = L_i- \left(\sum_{\ell: i\in T_\ell} \sum_{p\in RC_\ell}G_{\ell p} +  \sum_{\ell: i\in H_\ell} \sum_{p\in PC_\ell}G_{\ell p}\right)\quad &&\text{for}\quad i\in [1,n]. 
\end{aligned}
\end{equation}
where $G_{\ell p}$ and $L_i$ are defined as in \eqref{eq:nonbifunctional-hmm-general} and
\begin{equation}
\begin{aligned}
&\widetilde{G}_{\ell p} = -\left(\sum_{m\in R: \mathcal{E}_m=C_i}  \sum_{q\in RC_\ell\cup PC_\ell}G_{mq}\right).
\end{aligned}
\end{equation}
Similar to Section \ref{sec:ODEs}, here $G_{\ell p}$ and $\widetilde{G}_{\ell p}$ are the local and non-local parts of the ODEs for the compound $C_{\ell p}$. If $\widetilde{G}_{\ell p}=0$, $C_{\ell p}$ is not a bifunctional enzyme. 

\subsection{Analysis of the steady state equations}

Similar to Section \ref{sec:results}, we make the following assumption.
\begin{assumption}\label{assum:no_conservation_compounds_general}
There is no conservation law which contains only the compounds $\{c_{\ell p}: \ell\in R, ~ p\in RC_\ell\cup PC_\ell\}$.    
\end{assumption}
\begin{lemma}\label{lem:LFFtilde}
    The steady state equations are equivalent to
\begin{alignat*}{2}
G_{\ell p}&=0\quad &&\text{for}\quad \ell\in R, ~ p\in RC_\ell\cup PC_\ell, \mbox{ and }\\
L_i&= 0 \quad &&\text{for}\quad i\in[1,n]. 
\end{alignat*}
\end{lemma}
\begin{proof}
The argument is almost identical to that used to prove Lemma \ref{lem:Fij}.
\end{proof}

Next, we provide three lemmas which we will use in the proof of our main results for this section.
\begin{lemma}\label{lem:unimolecular}

Consider a reaction network $\mathcal{G}$ with $n$ species $\mathcal{S}=\{X_1, X_2, \dots, X_n\}$. Assume that set of complexes is $\mathcal{C}=\{0,X_1,\dots,X_n\}$ (i.e. at most unimolecular complexes). If $\mathcal{G}$ is weakly reversible and has a single connected component, then $\mathcal{G}$ has deficiency zero and thus its associated mass-action system $(\mathcal{G},\kappa)$ has a unique positive steady state for any choice of $\kappa$.
    
\end{lemma}

\begin{proof}
    It is easy to check that the stoichiometric subspace of $S$ contains all standard basis vectors of $\mathbb{R}^n$. Thus $\text{dim}(S)=n$ and its deficiency is $\delta = n+1 - 1 - n =0$. By Theorem \ref{thm:deficiencyzero}, $(\mathcal{G},\kappa)$ has a unique positive steady state for any choice of $\kappa$.
\end{proof}

\begin{lemma}\label{lem:scale_u}
Consider a reaction network $\mathcal{G}$ with $n$ species $\mathcal{S}=\{X_1, X_2, \dots, X_n\}$. Assume that set of complexes is $\mathcal{C}=\{0,X_1,\dots,X_n\}$ and that $G$ is weakly reversible and has a single connected component. 
For a choice of rate constants $\kappa$,  let the unique steady state of the mass-action system $(\mathcal{G},\kappa)$ be denoted by $x^*$. Let $u$ be a positive parameter, and consider a new choice of rate constants $\kappa_u$, defined as follows: 
\begin{enumerate}
\item If the reactant complex is $0$ then the rate constant of the reaction in $(\mathcal{G},\kappa)$ is multiplied by $u$ to obtain the rate constant of that reaction in $(\mathcal{G},\kappa_u)$. 
\item If the reactant complex is not $0$, then the rate constant of that reaction in $(\mathcal{G},\kappa_u)$ is the same as that in $(\mathcal{G},\kappa)$. 
\end{enumerate}

Then the positive steady state of $(\mathcal{G},\kappa_u)$ is $x^*_u = u\cdot x^*$.
\end{lemma}

\begin{proof}
Dividing the ODEs of the mass-action system $(\mathcal{G},\kappa_u)$ by $u$ and changing the variable $x\mapsto x/u$ produces the same ODEs as the mass-action system $(\mathcal{G},\kappa)$. Thus the positive steady state of $(\mathcal{G},\kappa_u)$ is $x^*_u = u\cdot x^*$.
\end{proof}

\begin{lemma}\label{lem:cip}
    For each hyperedge $\ell$ and each compound $c_{\ell p}: p\in RC_\ell$ in the reactant block, at steady state we must have
    \[
    c_{\ell p} = h_{\ell p} \left(\prod_{i\in T_\ell} s_i \right)\varepsilon_\ell,
    \]
    where $h_{\ell p}$ is a positive constant. This further implies
    \[
    c_{\ell p} = \frac{h_{\ell p}}{h_{\ell 1}}c_{\ell 1}.
    \]
\end{lemma}

\begin{proof}
For each hyperedge $\ell$, consider the reaction network $\mathcal{G}_\ell$ (and associated mass-action system $(\mathcal{G}_\ell,\kappa)$) that is obtained from its reactant block by doing the following: 
\begin{enumerate}
    \item replace the complex $\sum_{i\in T_\ell} S_i+\EE_\ell$ with the complex $0$, and 
    \item add a reaction from the terminal complex to the $0$ complex, with the same rate constant as the reaction from the terminal complex to the product block in the original network. 
\end{enumerate}
Assumption \ref{assum:blocks}(4) implies that $\mathcal{G}_\ell$ is weakly reversible and has a single connected component. Thus by Lemma \ref{lem:unimolecular}, there exists a unique positive steady state, in which the compound $C_{\ell p}$ has the concentration denoted by $h_{\ell p}$.

Returning to the full reaction network, from Lemma \ref{lem:LFFtilde}, the steady state satisfies $G_{\ell p}=0$ for all $p\in RC_\ell$. Let  $u=\left(\prod_{i\in T_\ell} s_i \right)\varepsilon_\ell$, then it is easy to see that the equations $G_{\ell p}=0$ for all $p\in RC_\ell$ are the same as the steady state equations of $(\mathcal{G}_\ell,\kappa_u)$ where $\kappa_u$ is defined as in Lemma \ref{lem:scale_u}. From Lemma \ref{lem:scale_u}, the steady state values of $c_{\ell p}$ satisfy
\[
c_{\ell p} = h_{\ell p} \cdot u = h_{\ell p} \left(\prod_{i\in T_\ell} s_i \right)\varepsilon_\ell.
\]
The second part of the lemma follows directly from this.
\end{proof}

Now we are ready to state the main results of this section on the consistency of substrate modification networks with general detailed model and conditions for ACR.

\begin{theorem}
Consider a substrate modification network whose detailed models satisfy Assumption \ref{assum:blocks}. Suppose there exists a positive current assignment on the skeleton $G$. Then the associated reaction network $\GG$ is consistent.
\end{theorem}
\begin{proof}
Suppose there exists a positive current assignment $\{I_\ell>0:\ell\in R\}$. We will show that there exists a choice of rate constants $\kappa$ such that the resulting mass-action system admits a positive steady state $x^*$ where the concentration of any species is equal to 1. Next, we show explicitly the choice of rate constants. From Lemma \ref{lem:LFFtilde}, we need to show that with this choice of rate constants, $x^*$ satisfies $L_i=0$ for all $i$ and $G_{\ell p}=0$ for all $\ell \in R, p\in RC_\ell\cup PC_\ell$. 
    \begin{itemize}
        \item First, we pick $k_{\ell} = I_{\ell}$ for all $\ell\in R$. At steady state $x^*$ where all $c_\ell=1$, we have
        \[
        L_i =  \sum_{\ell: i\in H_\ell} k_{\ell} c_{\ell} - \sum_{\ell: i\in T_\ell} k_{\ell}c_{\ell}  = \sum_{\ell: i\in H_\ell} I_{\ell}  - \sum_{\ell: i\in T_\ell} I_{\ell} = 0,
        \]
        where the last equality is due to the balance property of the currents.
        \item For the reactant block of hyperedge $\ell$, following the argument in the proof of Lemma \ref{lem:cip}, we consider a reaction network $\mathcal{G}_\ell$ (and associated mass-action system $(\mathcal{G}_\ell,\kappa)$) which is obtained by replacing the complex $\sum_{i\in T_\ell} S_i+\EE_\ell$ with the complex $0$. Again, since $\mathcal{G}_\ell$ is weakly reversible (and thus is consistent), there exists a choice of rate constants such that $(\mathcal{G}_\ell,\kappa)$ admits a unique steady state vector whose entries are all 1. This choice of rate constants ensures that $x^*$ satisfies $G_{\ell p}=0$ for all $\ell\in R, p\in RC_\ell$.
        \item For the product block of hyperedge $\ell$, we consider a reaction network $\mathcal{G'}_\ell$ (and associated mass-action system $(\mathcal{G'}_\ell,\kappa)$) which is obtained by replacing the complex $\sum_{i\in H_\ell} S_i+\EE_\ell$ with the complex $0$. Assumption \ref{assum:blocks} (1) ensures that $\mathcal{G'}_\ell$ is weakly reversible, and thus there exists a choice of rate constants such that $(\mathcal{G'}_\ell,\kappa)$ admits a unique steady state vector whose entries are all 1. This choice of rate constants ensures that $x^*$ satisfies $G_{\ell p}=0$ for all $\ell\in R, p\in PC_\ell$.
    \end{itemize}
Thus, $x^*$ is a positive steady state of the resulting mass-action system.
\end{proof}

The following definition is a general version of Definition \ref{def:bifunctional}. 

\begin{definition}\label{def:bifunctional_variation}
A pair of hyperedges $(\ell, m)$ is called a {\em bifunctionally linked pair} if a reactant compound of edge $\ell$ is the enzyme of edge $m$ (i.e. $\EE_{m} = C_{\ell p}$ for some $p\in RC_\ell$). Again, we call $\ell$  the c-edge and $m$ the e-edge of the bifunctionally linked pair. 
\end{definition}

\begin{theorem}\label{thm:ACR_general}
Consider a substrate modification network whose detailed models satisfy Assumption \ref{assum:blocks}. Suppose that the skeleton $G$ contains a pair of hyperedges $(\ell,m)$ that is both  bifunctionally linked and  current-matching. Then the associated reaction network $\GG$ has concentration robustness in the tail of $m$. 
Furthermore, if the tail of $m$ contains only one reactant substrate, then the associated reaction network $\GG$ has ACR in that substrate.
 \end{theorem}

\begin{proof}
From Lemma \ref{lem:LFFtilde}, at steady state, we have 
\[
L_i =  \sum_{\ell: i\in H_\ell} k_{\ell} c_{\ell 1} - \sum_{\ell: i\in T_\ell} k_{\ell}c_{\ell 1}  = 0
\]
for all $i\in [1,n]$. Thus, by setting $I_\ell = k_\ell c_{\ell 1}$, the set $\{I_\ell:\ell\in R\}$ forms a positive current assignment on $G$. Since $(\ell,m)$ is current-matching, their currents must equal for any choice of currents. Thus we have
\[
I_\ell = I_m \implies k_\ell c_{\ell 1}= k_m c_{m 1}\implies c_{\ell 1}= \frac{k_m}{k_\ell} c_{m 1}
\]
Suppose that the bifunctional enzyme is $C_{\ell p}$ for a fixed $p\in RC_\ell$. Then we have $\varepsilon_m=c_{\ell p}$. From Lemma \ref{lem:cip} we have
\[
c_{m1} = h_{m1}\left(\prod_{i\in T_m}s_{i}\right)\varepsilon_{m}= h_{m1}\left(\prod_{i\in T_m}s_{i}\right)c_{\ell p} = \left(\prod_{i\in T_m}s_{i}\right)h_{m1}\frac{h_{\ell p}}{h_{\ell 1}}c_{\ell 1}= \left(\prod_{i\in T_m}s_{i}\right) h_{m1}\frac{h_{\ell p}}{h_{\ell 1}}\frac{k_m}{k_\ell}c_{m1}.
\]
Thus we have 
\[
\prod_{i\in T_m}s_{i} = \frac{h_{\ell 1}k_\ell}{h_{m1}h_{\ell p}k_{m}},
\]
at any positive steady state. Thus $\GG$ has concentration robustness in the tail of $m$. Furthermore, if there is only one $i\in T_m$, then we must have
\[
s_i = \frac{h_{\ell 1}k_\ell}{h_{m1}h_{\ell p}k_{m}}
\]
and thus $\GG$ has ACR in substrate $S_i$, the only reactant substrate of the e-edge $m$.
\end{proof}
\begin{remark}
It is important to note from Theorem \ref{thm:ACR_general} that if a {\em product compound} is a bifunctional enzyme, we may {\em not} have an ACR species. To illustrate this, consider a simple substrate modification network $S\rightleftarrows S^*$ with the following detailed models:
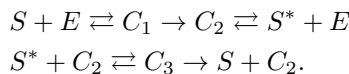
\begin{align*}
  &S+E\rightleftarrows C_{1} \to C_{2}\rightleftarrows S^*+E \\ 
  &S^*+C_{2}\rightleftarrows C_3 \to S+C_2.  
\end{align*}
In this case, a product compound, $C_2$, is a bifunctional enzyme. It can be checked (for example by Mathematica) that the associated mass-action system does not have an ACR species.
\end{remark}

Finally, similar to Remark \ref{remark:non-mass-action}, the results in Theorem \ref{thm:ACR_general} can be extended to more general kinetics beyond mass-action kinetics.

\bibliographystyle{unsrt}
\bibliography{acr}

\section*{Acknowledgments}
B.J. was supported by NSF grant DMS-2051498. 

\section*{Author contributions}

B.J. and T.D.N. both contributed to the conceptualization, analytical methodology, formal analysis, and the writing of the paper.

\section*{Competing interests}
The authors declare no competing interests.

\end{document}